\documentclass[twoside]{article}

\usepackage{color}
\usepackage{amssymb}
\usepackage{dsfont}
\usepackage{amssymb,amsmath,ulem,cancel}

\numberwithin{equation}{section}

\newtheorem{theorem}{Theorem}[section]
\newtheorem{lem}{Lemma}[section]
\newtheorem{pro}{Proposition}[section]
\newtheorem{cor}{Corollary}[section]
\newtheorem{rem}{Remark}[section]
\newtheorem{rems}{Remarks}[section]
\newtheorem{ex}{Example}[section]
\newtheorem{defi}{Definition}[section]
\newtheorem{hyp}{Assumption}[section]
\newtheorem{con}{Conjecture}[section]

\newcommand{\bt}{\begin{theorem}}
\newcommand{\et}{\end{theorem}}
\newcommand{\bl}{\begin{lem}}
\newcommand{\el}{\end{lem}}
\newcommand{\bp}{\begin{pro}}
\newcommand{\ep}{\end{pro}}
\newcommand{\bcor}{\begin{cor}}
\newcommand{\ecor}{\end{cor}}
\newcommand{\bcon }{\begin{con} \rm }
\newcommand{\econ }{\end{con}}
\newcommand{\lab }{\label }
\newcommand{\bd}{\begin{defi} \rm }
\newcommand{\ed}{\end{defi}}
\newcommand{\brem }{\begin{rem} \rm }
\newcommand{\erem }{\end{rem}}
\newcommand{\brems }{\begin{rems} \rm }
\newcommand{\erems }{\end{rems}}
\newcommand{\bhyp }{\begin{hyp} \rm }
\newcommand{\ehyp }{\end{hyp}}
\newcommand{\bex}{\begin{ex} \rm }
\newcommand{\eex}{\end{ex}}
\newcommand{\be}{\begin{equation}}
\newcommand{\ee}{\end{equation}}
\newcommand{\bde}{\begin{displaymath}}
\newcommand{\ede}{\end{displaymath}}
\newcommand{\beq}{\begin{eqnarray*}}
\newcommand{\eeq}{\end{eqnarray*}}
\newcommand{\beqa}{\begin{eqnarray}}
\newcommand{\eeqa}{\end{eqnarray}}
\newcommand{\bea}{\begin{align*}}
\newcommand{\eea}{\end{align*}}

\def\I{\mathds{1}}
\def\wh{\widehat}
\def\wt{\widetilde}

\def\phi{\varphi }

\newcommand{\Cnq}{\widehat{q}}
\newcommand{\Classaq}{{\cal A}({\mathbb Q})}

\newcommand{\Classaptb }{{\cal A}(\widetilde{\mathbb P}^{\beta })}
\newcommand{\Classapl }{{\cal A}(\widetilde{\mathbb P}^l)}

\newcommand{\aass}{\mbox{\rm a.s.}}
\newcommand{\aaee}{\mbox{\rm a.e.}}

\newcommand{\ssx}{s}

\newcommand{\Leb }{\ell }

\newcommand{\pA}{A}
\newcommand{\pC}{C}

\newcommand{\pCc}{c}

\newcommand{\etab}{\eta^b}
\newcommand{\etal}{\eta^l}

\newcommand{\wtWl}{\widetilde{W}^l}
\newcommand{\wtWb}{\widetilde{W}^{\beta }}

\newcommand{\Blr}{B^{l}}
\newcommand{\Bbr}{B^{b}}

\newcommand{\Bilr}{B^{i,l}}
\newcommand{\Bibr}{B^{i,b}}

\newcommand{\Prir}{P^r}
\newcommand{\rir}{r}
\newcommand{\rll}{r^{l}}
\newcommand{\rbb}{r^{b}}

\def\t1{\tau_{(1)}}

\def\rr{\mathbb R}
\def\ff{{\mathbb F}}

\def\gg{{\mathbb G}}

\def\G{{\cal G}}

\def\VLL{V^0}
\def\P{\mathbb P}

\def\PT{\wt {\mathbb P}}
\def\PTb{\wt {\mathbb P}^{\beta } }

\def\Q{\mathbb Q}

\def\EP{{\mathbb E}_{\mathbb P}}

\def\C-FVA{{\rm C-FVA}}

\newcommand{\wHzero}{\wh{\mathcal{H}}_{0}^{2}}
\newcommand{\wHzerd}{\wh{\mathcal{H}}_{0}^{2,d}}

\newcommand{\sumik}{\textstyle{\sum}}

\newcommand{\Keywords}[1]{\par\noindent{\small{\bf Keywords\/}: #1}}
\newcommand{\Class}[1]{\par\noindent{\small{\bf Mathematics Subjects Classification (2010)\/}: #1}}

\title{{\Large \bf A BSDE APPROACH TO FAIR BILATERAL PRICING \\ \vskip 1 pt UNDER ENDOGENOUS COLLATERALIZATION} \vskip 65 pt }

\author{Tianyang Nie and Marek Rutkowski\footnote{The research of Tianyang Nie and Marek Rutkowski
was supported under Australian Research Council's Discovery Projects funding scheme (DP120100895).}
\\ School of Mathematics and Statistics \\ University of Sydney
\\ Sydney, NSW 2006, Australia}

\date{\vskip 45 pt 1 December 2014 \vskip 45 pt}

\begin{document}

\maketitle

\begin{abstract}
Results from Nie and Rutkowski \cite{NR2,NR4} are extended to the case of the margin account, which may depend on the contract's value for the hedger and/or the counterparty (recall that the collateral was given exogenously in \cite{NR2,NR4}). The present work generalizes also the papers by Bergman \cite{B-1995}, Mercurio \cite{M-2013} and Piterbarg \cite{P10}. Using the comparison theorems for BSDEs, we derive inequalities for the unilateral prices and we give the range for its fair bilateral prices. We also establish results yielding the link to the market model  with a single interest rate. In the case where the collateral amount is negotiated between the counterparties, so that it depends on their respective unilateral values, the backward stochastic viability property studied by Buckdahn et al. \cite{BQR-2000} is used to derive the bounds on fair bilateral prices.
\vskip 20 pt
\vskip 20 pt
\Class{91G20,$\,$91G80}
\end{abstract}


\newpage

\section{Introduction} \label{sec1}

In Bielecki and Rutkowski \cite{BR-2014}, the authors introduced a generic nonlinear market model which includes several risky assets, multiple funding accounts, as well as the margin account for collateral (for related studies by other authors, see also \cite{BCPP11,BK09,BK11,SC12a,SC12b,PPB12,P10}). We continue their study by examining the pricing and hedging of a derivative contract from the perspective of the hedger and his counterparty. Since we work within a nonlinear trading framework, the prices computed by the two parties of
 a contract do not necessarily coincide and thus our goal is to compare these prices and to derive the range for no-arbitrage bilateral prices.  As emphasized in \cite{BR-2014,NR2,NR3}, the initial endowments of the hedger and the counterparty become important factors in arbitrage pricing in the nonlinear setup. In \cite{NR2,NR4}, we studied collateralized contracts in the model with partial netting and Bergman's model, respectively. Using comparison theorems for BSDEs, we derived the range for either fair bilateral prices or bilaterally profitable prices. It should be stressed that in \cite{NR2,NR4}, the collateral amount was assumed to be exogenously specified and thus it was  independent of unilateral values of the contract for the two parties. By contrast, we study here a more realistic situation where the collateral is endogenous, meaning that it may depend on the marked-to-market value of the contract either for one party (say, the hedger) or it is negotiated between them. Although we focus here on two particular instances of market models, it is clear that the method
  developed in this work can be applied to a large variety of models and/or collateral covenants.

Motivated by the seminal paper by Bergman \cite{B-1995}, we first consider an extension of his trading model to the case of endogenous collateral. To the best of our knowledge, the case of endogenous collateral was not studied in the existing literature, except for the special case of the proportional collateral examined by Piterbarg \cite{P10} and Mercurio \cite{M-2013}. We give here essential extensions of their results using the BSDE approach. First, we consider general collateralized contracts, rather only than path-independent European claims. Second, in \cite{M-2013}, the collateralization of the hedger (resp., the counterparty) was
postulated to be a constant proportion of the hedger's (resp., the counterparty's) value, which apparently means that the two parties
either post or receive the collateral amounts specified by two different margin accounts. This is clearly inconsistent
with the market practice where the collateral amount posted by one party coincides with the amount received by another party.
We derive inequalities satisfied by unilateral prices of a contract and we give the range for its fair bilateral prices.
We show that if the collateralization depends on the values for the hedger and the counterparty, the backward stochastic viability property  (BSVP) plays an important role in derivation of pricing inequalities. Motivated by results from the papers by Buckdahn et al. \cite{BQR-2000} and Hu and Peng \cite{HP-2006}, we obtain the range of fair bilateral prices for European contingent claims.  In the second step, we consider the market model with partial netting under the assumption of full rehypothecation of the cash collateral. Once again, in contrast with our previous work \cite{NR2}, we study here the case of the collateral depending on the hedger's value and/or counterparty's value. We establish similar results as for Bergman's model.  It is worth noting, however, that the model with partial netting enjoys some additional properties with respect to the class of monotone contracts, which are not necessarily shared by Bergman's model. This emphasizes the impact of asset-specific funding costs on properties of hedging strategies and prices of contracts.

The work is organized as follows. In Section \ref{sec2}, we recall some definitions and assumptions introduced in our previous works \cite{BR-2014,NR2,NR4}. In Section \ref{sec3} and \ref{sec4}, we examine extensions of the model studied by
Bergman \cite{B-1995} and Mercurio \cite{M-2013}. In Section \ref{sec3}, we consider the case where the collateral depends only on the hedger's value and we establish inequalities for unilateral prices of a general contract. Moreover, we extend the results from \cite{M-2013} regarding the relationship to the market model with a single uncertain interest rate. In Section \ref{sec4}, we study the case where the collateral depends on both the hedger's and the counterparty's values under the assumption that the risky asset is driven by a Brownian motion. Using the BSVP technique from \cite{BQR-2000}, we derive the inequalities for unilateral prices. In Sections \ref{sec5} and \ref{sec6}, we examine the model with partial netting and we obtain similar results for the range of fair bilateral prices.  We also show that the model with partial netting has some additional properties of independence of the initial endowment and/or positive homogeneity with respect to particular classes of contracts. We thus conclude that no unified theory can be developed in the non-linear framework, so that each particular setup should be studied on a standalone basis.

\newpage

\section{Preliminaries} \lab{sec2}

We provide here a very brief summary of concepts and notation introduced in  \cite{BR-2014,NR2,NR4}.
For more details and explanations, the reader is referred to the original papers.
Let $T>0$ be a fixed finite trading horizon date for our model of the financial market.
 We denote by $(\Omega, \G, \gg , \P)$ a filtered probability space satisfying the usual conditions of right-continuity and completeness, where the filtration $\gg = (\G_t)_{t \in [0,T]}$ models the flow of information available to all traders. For convenience, we assume that the initial $\sigma$-field ${\cal G}_0$ is trivial. Moreover, all processes introduced in what follows are implicitly assumed to be $\gg$-adapted and any semimartingale is assumed to be c\`adl\`ag.
As in \cite{NR2,NR4}, for any $i=1,2, \dots, d$, we use the following notation for the market data:
\hfill \break $A$ --  a {\it bilateral financial contract}, or simply a {\it contract}. The process $A$ is finite variation  and it represents the {\it cumulative cash flows} of a given contract from time 0 till its maturity date $T$,
\hfill \break $C$ -- the cash collateral, specified as a $\gg$-adapted process satisfying $C_T=0$,
 \hfill \break
$S^i$ -- the {\it ex-dividend price} of the $i$th risky asset with the {\it cumulative dividend stream} $\pA^i$,
\hfill \break $B^l$ (resp., $B^b$) --  the {\it lending} (resp., {\it borrowing}) {\it cash account},
\hfill \break $\Bilr$ (resp., $\Bibr$) -- the {\it lending} (resp., {\it borrowing}) {\it funding account} associated with the $i$th risky asset,
\hfill \break $B^c$ -- the process specifying the interest paid/received on the collateral account received/posted.

\bhyp \lab{assumption for primary assets}
We work throughout under the following assumptions: \hfill \break
(i) $S^i$ is a semimartingale and $\pA^i$ is a process of finite variation with $\pA^i_{0}=0$.\hfill \break
(ii) the processes $B^{l}, B^{b},B^{i,l},B^{i,b}$ and $B^c$ are strictly positive, continuous processes of finite variation with $B^l_0=B^b_0=B^{i,l}_0=B^{i,b}_0=B^c_0=1$.\hfill \break
(iii) in the case of a model with partial netting, we also assume that $B^{i,l}=B^l$ for every $i$, \hfill \break
(iv) $dB^{l}_{t}=r^{l}_{t}B^{l}_{t}\, dt$, $dB^{b}_{t}=r^{b}_{t}B^{b}_{t}\, dt$, $dB^{i,b}_{t}=r^{i,b}_{t}B^{i,b}_{t}\, dt$ and $dB^{c}_{t}=r^{c}_{t}B^{c}_{t}\, dt$,  for some $\mathbb{G}$-adapted and bounded processes $r^{l}$, $r^{b}$, $r^{i,b}$ and $r^c$. Moreover, we postulate that $0 \leq \rll \leq \rbb$ and $\rll\leq r^{i,b}$.
\ehyp

We define the interest process of the margin account by setting $F^{C}_t := -\int_0^t  r^c_u \pC_u \, du$ for every~$t$
and we denote $A^{C}:=A+C+F^{C}$. For a {\it collateralized hedger's trading strategy} $(x,\phi , \pA , \pC )$, we write: \hfill \break
(i) $V_t(x,\phi , \pA ,\pC)$ -- the hedger's wealth at time $t$,
\hfill \break (ii) $V_t^p(x,\phi , \pA ,\pC)$ -- the value of hedger's portfolio at time $t$.

Note that $V^p_t(x,\phi , \pA ,\pC) - V_t(x,\phi , \pA ,\pC)= C_t$ measures the impact of the margin account represented by the collateral amount $C_t$ on the hedger's trading strategy under the standing assumption of {\it full rehypothecation}.
Finally, we set $V_{t}^{0}(x):=xB_{T}^l \I_{\{x\ge0\}}+x\Bbr_T\I_{\{x<0\}}$ where $x=x_1$ (resp., $x=x_2$) is the {\it initial endowment} of the hedger (resp., the counterparty) at time 0.

\bd \label{definition of ex-dividend price}
Any ${\cal G}_{t}$-measurable random variable for which a replicating strategy for $A$ over $[t,T]$ exists is called the {\it hedger's ex-dividend price} at time $t$ for a contract $(A,C)$ and it is denoted by $P^{h}_t(x_1, A, C)$, so that for some
self-financing trading strategy $\phi $, which replicates $(A,C)$, we have
\bde
V_T(\VLL_{t}(x_1)+P^{h}_t(x_1, A, C), \phi , A -A_t, C) = \VLL_T (x_1).
\ede
For an arbitrary level $x_2$ of the counterparty's initial endowment and a strategy $\widetilde{\phi}$ replicating $(-A,-C)$,
the {\it counterparty's ex-dividend price} $P^{c}_t(x_2, -A, -C)$ at time $t$ for a contract $(-A,-C)$ is implicitly given by the equality
\bde
V_T(\VLL_{t}(x_2)-P^{c}_t(x_2, -A, -C), \widetilde{\phi }, -A+A_t, -C) = \VLL_T (x_2).
\ede
\ed

By a {\it fair bilateral price}, we mean any level of the price at which no arbitrage opportunity arises for either of the
two  parties. Hence the range of fair bilateral prices at time $t$ is formally defined as follows.

\bd  \label{range of fair}
The  $\G_t$-measurable interval
\bde
{\cal R}^f_t (x_1,x_2) := \big[ P^{c}_t(x_{2},-A,-C), P^{h}_t (x_{1},A,C) \big]
\ede
is called the {\it range of fair bilateral prices} at time $t$ of an OTC contract $(A,C)$ between the hedger and the counterparty.
\ed

\section{Bergman's Model with Hedger's Collateral} \label{sec3}

In Sections \ref{sec3} and \ref{sec4}, we consider an extended version of the model studied by Bergman \cite{B-1995}. For a detailed analysis of this model, we refer to the recent work by Nie and Rutkowski \cite{NR4}. Note that in this framework the funding accounts $B^{i,l}$ and $B^{i,b}$ are not introduced.

Following \cite{BR-2014,NR2,NR4}, we introduce the auxiliary processes $\wt S^{i,l,\textrm{cld}}$ and $\wt S^{i,b,\textrm{cld}}$, which are given by the following expressions, for $i=1,2, \dots , d$,
\bde
\wt S^{i,l,{\textrm{cld}}}_t :=  (\Blr_t)^{-1}S^i_t + \int_{(0,t]} (\Blr_u)^{-1} \, d\pA^i_u
\ede
and
\bde
\wt
S^{i,b,{\textrm{cld}}}_t :=  (\Bbr_t)^{-1}S^i_t + \int_{(0,t]} (\Bbr_u)^{-1} \, d\pA^i_u .
\ede
It is easy to see that the dynamics of these processes are
\be \label{cumulative dividend risk asset price1}
d\wt S^{i,l,{\textrm{cld}}}_t=(\Blr_t)^{-1}\big(dS^i_t + d\pA^i_t - \rll_t S^i_t \, dt \big)
\ee
and
\be  \label{cumulative dividend risk asset price2}
d\wt S^{i,b,{\textrm{cld}}}_t=(\Bbr_t)^{-1}\big(dS^i_t + d\pA^i_t - \rbb_t S^i_t \, dt \big).
\ee
As in  \cite{NR4}, we consider an arbitrary self-financing trading strategy $\varphi=(\xi^1, \dots , \xi^d ,\varphi^l,\varphi^b,\etal , \etab )$ where $\etal_t =(B^{\pCc,l}_t)^{-1} \pC_t^-$ and $\etab_t =-  (B^{\pCc,b}_t)^{-1}\pC_t^+$.
Since we assume here that $B^{\pCc,l}_t=B^{\pCc,b}_t=B^{\pCc}_t$, a trading strategy $\varphi$ reduces to $(\xi^1, \dots , \xi^d ,\varphi^l,\varphi^b,\eta)$ where $\eta_t=-(B^{\pCc}_t)^{-1}\pC_t$. Let us denote
\bde
A^{C,l}_t := \int_{(0,t]}(\Blr_{u})^{-1}\, dA^C_{u}, \quad A^{C,b}_t := \int_{(0,t]}(\Bbr_{u})^{-1}\, dA^C_{u}.
\ede
From Proposition 2.9 in \cite{NR4}, it is known that the process $Y^{l}:=(B^{l})^{-1}V^p(x,\phi , \pA ,\pC)$ satisfies
\be  \label{Bergman model lending BSDE}
dY_{t}^{l} = \sum_{i=1}^dZ^{l,i}_t \, d\wt S^{i,l,{\textrm{cld}}}_t+G_{l}(t,Y_{t}^{l},Z_{t}^{l})\, dt+ d\pA_t^{C,l}
\ee
where $Z^{l,i}:=\xi^i ,\, i=1,2,\ldots,d$ and the generator $G_l$ is given by the following expression, for all $(\omega , t,y,z)\in \Omega \times [0,T] \times \mathbb{R}\times\mathbb{R}^{d}$,
\bde
G_{l}(t,y,z)=\sum_{i=1}^d r_{t}^{l}(B_{t}^{l})^{-1} z^i S^i_t+(B_{t}^{l})^{-1}\bigg(r_{t}^{l}\Big(yB_{t}^{l}-\sum_{i=1}^d z^i S^i_t\Big)^+-r_{t}^{b}\Big(yB_{t}^{l}-  \sum_{i=1}^d z^i S^i_t\Big)^- \bigg)-r_{t}^{l}y.
\ede
Similarly, the process $Y^{b}:=(B^{b})^{-1}V^p (x,\phi , \pA ,\pC)$ is governed by
\be \label{Bergman model borrowing BSDE}
dY_{t}^{b} = \sum_{i=1}^dZ^{b,i}_t \, d\wt S^{i,b,{\textrm{cld}}}_t+G_{b}(t,Y_{t}^{b},Z_{t}^{b})\, dt+ d\pA_t^{C,b}
\ee
where  $Z^{b,i}=\xi^i,\, i=1,2,\ldots,d$ and the generator $G_b$ equals, for all $(\omega ,t,y,z)\in \Omega \times [0,T] \times \mathbb{R}\times\mathbb{R}^{d}$,
\bde
G_{b}(t,y,z)=\sum_{i=1}^d r_{t}^{b}(B_{t}^{b})^{-1}z^i S^i_t+(B_{t}^{b})^{-1}\bigg(r_{t}^{l}\Big(yB_{t}^{b}-\sum_{i=1}^d z^i S^i_t\Big)^+-r_{t}^{b}\Big(yB_{t}^{b}-  \sum_{i=1}^d z^i S^i_t\Big)^- \bigg)-r_{t}^{b}y.
\ede
Recall that the initial endowment of the hedger (resp., the counterparty) is denoted by $x_{1}$ (resp., $x_{2}$). Without loss of generality, we assume throughout that $x_{1}\ge0$ and we consider an arbitrary level of $x_2$.
Furthermore, in Sections \ref{sec3} and \ref{sec5}, we work under the following  standing assumption of {\it hedger's collateral}, that is,
the situation where the collateral amount only depends on the hedger's wealth $V^{h}:=V(x_{1},\varphi,A,C)$.

\bhyp \label{hypo1}
{\rm The {\it hedger's collateral} $C$ is given by the equality
\be \label{collateral 1}
C_t=q(V_{t}^{0}(x_1)-V^{h}_t)
\ee
for some uniformly Lipschitz continuous function $q:\mathbb{R}\rightarrow\mathbb{R}$ such that $q(0)=0$.}
\ehyp

\bex \label{ex1}
For instance, the hedger's collateral $C$ can be specified as in \cite{BR-2014} (see equation (4.10) therein) through the following expression
\bde
C_{t}=(1+\alpha_1)\big( V_{t}^{0}(x)-V^h_{t}\big)^{+}
-  (1+\alpha_2)\big( V_{t}^{0}(x)-V^h_{t} \big)^{-}
\ede
for some bounded {\it haircut} processes $\alpha_1, \alpha_2$, so that $q(y)=(1+\alpha_1)y^{+}-(1+\alpha_2)y^{-}$.
It is clear that $q$ is uniformly Lipschitz continuous and $q(0)=0$.
The case of the fully collateralized contract, from the perspective of the hedger, is obtained by taking $q(y)=y$, that is,
by setting $\alpha_1=\alpha_2=0$.
\eex

\subsection{Initial Endowments of Equal Signs} \label{sec3.1}

We first examine the case where the initial endowments are of the same sign, specifically, we assume that $x_{1}\ge0$ and $x_{2}\ge0$. The next assumption postulates the existence of a `martingale measure' in the present setup. All probability measures are assumed to be defined
on $(\Omega , {\cal G}_T)$.

\bhyp \label{assumption for lending cumulative dividend price}
There exists a probability measure $\PT^l $ equivalent to $\P $ such that the processes $\wt S^{i,l,\textrm{cld}},\, i=1,2, \dots ,d$ are $(\PT^l , \gg)$-local martingales.
\ehyp

The following result is borrowed from \cite{NR4} (see Proposition 2.1 therein). Let us stress that the arbitrage-free property is here understood in the sense of \cite{BR-2014} (see Section 3.1 therein).

\bp \label{Bergman model proposition for arbitrage free}
If the initial endowments satisfy $x_1 \geq 0 ,\, x_2 \geq 0$ and Assumption \ref{assumption for lending cumulative dividend price} is valid, then Bergman's model is arbitrage-free  for the hedger and the counterparty with respect to any contract $(A,C)$.
\ep

In order to address the issue of bilateral pricing using a BSDE approach, we need to impose additional assumptions
on the dynamics of risky assets.
We will work under the following assumption regarding the quadratic variation process for continuous martingales $\wt S^{l,\textrm{cld}}$. Note that $^{\ast}$ stands for the transposition and, as in \cite{NR2,NR4}, we define
the matrix-valued process $\mathbb{S}$ given by
\[
\mathbb{S}_{t}:=
\begin{pmatrix}
S^{1}_{t} & 0 & \ldots & 0 \\
0 & S^{2}_{t} & \ldots & 0 \\
\vdots & \vdots & \ddots & \vdots\\
0 & 0 & \ldots & S^{d}_{t}
\end{pmatrix}.
\]

\bhyp \label{changed assumption for lending cumulative dividend price}
We postulate that: \hfill \break
(i) there exists a probability measure $\PT^l $ equivalent to $\P $ such that $\wt S^{l,\textrm{cld}}$ is a continuous, square-integrable, $(\PT^l , \gg)$-martingale and has the predictable representation property with respect to the filtration $\gg$ under~$\PT^l$, \hfill \break
(ii)  there exists an $\mathbb{R}^{d\times d}$-valued, $\gg$-adapted process $m^{l}$ such that
\be \label{vfvf1}
\langle \wt S^{l,\textrm{cld}}\rangle_{t}=\int_{0}^{t}m^{l}_{u}(m_{u}^{l})^{\ast}\,du
\ee
where the process $m^{l}(m^{l})^{\ast}$ is invertible and satisfies $m^{l}(m^{l})^{\ast}=\mathbb{S}\sigma\sigma^{\ast}\mathbb{S}$.  Here $\sigma$ is a $d$-dimensional square matrix of $\gg$-adapted processes satisfying the {\it ellipticity condition}: there exists a constant $\Lambda>0$
\be \label{elli}
\sum_{i,j=1}^{d} \left( \sigma_{t}\sigma^{\ast}_{t}\right)_{ij} a_{i}a_{j} \ge \Lambda |a|^{2}
 = \Lambda a^{\ast}a, \quad \forall \, a \in \mathbb{R}^{d},\, t \in[0,T].
\ee
\ehyp

Following Nie and Rutkowski \cite{NR3}, but with $Q_t=t$, we denote by $\wHzerd $ the subspace of all $\mathbb{R}^{d}$-valued, $\gg$-adapted processes $X$ such that
\be \label{defhh}
|X|_{\wHzerd}^{2} := \EP \bigg[ \int_{0}^{T} | X_{t}|^{2} \,dt \bigg] < \infty
\ee
and, for brevity, we write $\widehat{\mathcal{H}}_0^2 := \widehat{\mathcal{H}}_0^{2,1}$. Also, let $\widehat{L}^{2}_{0}$ stand
for the space of all real-valued, $\mathcal{G}_{T}$-measurable random variables $\eta$ such that $|\eta|_{\widehat{L}^2_{0}}^{2}
= \EP (\eta^{2})<\infty $.

\bd \label{defadmi}
For any probability measure $\Q$, we denote by $\Classaq $ the following class of a real-valued, $\mathbb{G}$-adapted processes
$\Classaq := \big\{ X \in\widehat{\mathcal{H}}_{0}^{2} \mbox{\ and\ } X_{T}\in \widehat{L}_{0}^{2} \mbox{\ under\  } \mathbb{Q} \big\}$.
\ed

Definition \ref{defadmi} will serve to define the class of {\it admissible} contracts with the choice of $\Q$ depending on a particular setup at hand. Let us stress that for any contract $(A,C)$ the statement that $A \in \Classaq$ will mean that the process $A-A_0$ of future cash flows belongs to the class $\Classaq$. Recall that the initial cash flow $A_0$ of a contract $(A,C)$ represents its initial price, so that is not given a priori.

For the reader's convenience, we first recall a result concerning the case of an exogenous collateral $C$ (see Propositions 3.1 and 3.2 in \cite{NR4}, as well as Proposition 5.2 in \cite{BR-2014}).

\bp \label{pro3.2}
Let $x_1 \geq 0,\, x_2\ge0$ and Assumptions \ref{hypo1} and \ref{changed assumption for lending cumulative dividend price} be valid. Then for any contract $(A,C)$ such that $A^{C,l} \in \Classapl $, the hedger's ex-dividend price equals $P^{h}(x_{1},A,C) =  \Blr (Y^{h,l,x_{1}} - x_{1})-C$ where the pair $(Y^{h,l,x_{1}}, Z^{h,l,x_{1}})$ is the unique solution to the BSDE
\begin{equation} \label{Bergman model BSDE with positive x for hedger}
\left\{ \begin{array}
[c]{l}
dY^{h,l,x_{1}}_t = Z^{h,l,x_{1},\ast}_t \, d \wt S^{l,{\textrm{cld}}}_t
+G_l \big(t, Y^{h,l,x_{1}}_t, Z^{h,l,x_{1}}_t \big)\, dt + dA^{C,l}_t, \medskip\\
Y^{h,l,x_{1}}_T=x_{1},
\end{array} \right.
\end{equation}
and the counterparty's ex-dividend price equals $P^{c} (x_2,-A,-C) =-\Blr (Y^{c,l,x_2} - x_2)+C$
where the pair $(Y^{c,l,x_2}, Z^{c,l,x_2})$ is the unique solutions to the BSDE
\be \label{Bergman model BSDE with positive x for counterparty}
\left\{ \begin{array}
[c]{l}
dY^{c,l,x_2}_t = Z^{c,l,x_2,\ast}_t \, d \wt S^{l,{\textrm{cld}}}_t
+G_l \big(t, Y^{c,l,x_2}_t, Z^{c,l,x_2}_t \big)\, dt - dA^{C,l}_t, \medskip\\
Y^{c,l,x_2}_T=x_2 .\end{array}
\right.
\ee
\ep

In the next result, the hedger's collateral $C$ is given by equation \eqref{collateral 1}. Note that the generator $g_{l}$ depends explicitly on the process $Y^1$, which in turn is defined as a part of the solution of BSDE (\ref{main BSDE hedger}). This means that the counterparty's BSDE (\ref{main BSDE counterparty}) is coupled with the hedger's BSDE (\ref{main BSDE hedger}). It is thus crucial to note that the hedger's price $P^{h}(x_{1},A,C)$ depends only on his initial endowment $x_1$. By contrast, the counterparty's price depends on both initial endowments, $x_1$ and $x_2$, so that it would be suitable to denote it as $P^{c}(x_{1},x_2,-A,-C)$. However, for ease of notation, we shall write  $P^{c}(x_2,-A,-C)$, while keeping in mind that this process depends on $x_1$ as well.

\bp \label{hedger collateral dependent Bergman model ex-dividend price}
Let $x_1 \geq 0,\, x_2\ge0$ and Assumptions \ref{hypo1} and \ref{changed assumption for lending cumulative dividend price} be valid. Then for any contract $(A,C)$ such that $A \in \Classapl $, the hedger's ex-dividend price equals $P^h:=P^{h}(x_{1},A,C) = Y^1$ where $(Y^{1}, Z^{1})$ is the unique solution to the BSDE
\be \label{main BSDE hedger}
\left\{ \begin{array}
[c]{l}
dY_t^1 = Z^{1,\ast}_t \, d \wt S^{l,{\textrm{cld}}}_t
+f_l \big(t, x_1,Y^1_t, Z^1_t \big)\, dt + dA_t, \medskip\\
Y^1_T=0,
\end{array} \right.
\ee
with the generator $f_l$ given by
\begin{align*}
&f_{l}(t,x_1,y,z) =r_{t}^{l}(B_t^{l})^{-1}z^\ast S_t -x_1 B_{t}^{l}r_{t}^{l}-r_{t}^{c}q(-y) \\
&\mbox{}+r_{t}^{l}\Big(y+q(-y)+x_1B_{t}^{l}-(B_{t}^{l})^{-1}z^\ast S_t\Big)^+
-r_{t}^{b}\Big(y+q(-y)+x_1B_{t}^{l}-(B_{t}^{l})^{-1}z^\ast S_t\Big)^-
\end{align*}
and the counterparty's ex-dividend price equals $P^{c}:=P^{c} (x_2,-A,-C) =Y^2$ where $(Y^{2}, Z^{2})$ is the unique solution to the BSDE
\be \label{main BSDE counterparty}
\left\{ \begin{array}
[c]{l}
dY_t^2 = Z^{2,\ast}_t \, d \wt S^{l,{\textrm{cld}}}_t
+g_l \big(t, x_2,Y^2_t, Z^2_t \big)\, dt + dA_t, \medskip\\
Y^2_T=0,
\end{array} \right.
\ee
with the generator $g_l$ given by
\begin{align*}
&g_{l}(t,x_2,y,z) =r_{t}^{l}(B_t^{l})^{-1}z^\ast S_t +x_2B_{t}^{l}r_{t}^{l}-r_{t}^{c}q(-Y^1_{t}) \\
&\mbox{}-r_{t}^{l}\Big(-y-q(-Y^1_{t})+x_2B_{t}^{l}+(B_{t}^{l})^{-1}z^\ast S_t\Big)^+
+r_{t}^{b}\Big(-y-q(-Y^1_{t})+x_2B_{t}^{l}+(B_{t}^{l})^{-1}z^\ast S_t\Big)^-.
\end{align*}
\ep

\begin{proof}
Since the collateral amount is not exogenously specified in the present framework, the process $C$ may depend on the hedger's value, and thus Proposition \ref{pro3.2} does not cover the current situation. However, from the proof of Proposition 5.2 in \cite{BR-2014},  one can deduce that if BSDEs (\ref{Bergman model BSDE with positive x for hedger}) and (\ref{Bergman model BSDE with positive x for counterparty}) have a unique solution, then the relationships $P^{h}(x,A,C) =  \Blr (Y^{h,l,x_1} - x_1)-C$ and $P^{c} (x_2,-A,-C) =- \Blr (Y^{c,l,x_2} - x_2)+C$ are still valid.

It is also worth stressing that we cannot apply directly the results of \cite{NR3} to solve BSDEs (\ref{Bergman model BSDE with positive x for hedger}) and (\ref{Bergman model BSDE with positive x for counterparty}), since the process $A^C$ depends also on $Y^{h,l,x}$. However, since $P^{h}:=  \Blr (Y^{h,l,x_1} - x_1)-C$, we have that $P^{h}_T=0$ and thus, by letting $\widetilde{Z}^{h,l,x_1}:=B^{l}Z^{h,l,x_1}$, we obtain
\begin{align*}
dP^{h}_t& =B_{t}^{l}Z^{h,l,x_1,\ast}_t \, d \wt S^{l,{\textrm{cld}}}_t
+B_{t}^{l}G_l \big(t, Y^{h,l,x_1}_t, Z^{h,l,x_1}_t \big)\, dt + r_{t}^{l} B_{t}^{l}(Y^{h,l,x_1} - x)\, dt+ dA^C_t-dC_{t}\\
&=B_{t}^{l}Z^{h,l,x_1,\ast}_t \, d \wt S^{l,{\textrm{cld}}}_t
+r_{t}^{l}Z^{h,l,x_1,\ast}_tS_t\, dt+r_{t}^{l}\Big(Y^{h,l,x_1}_{t}B_{t}^{l}-Z^{h,l,x_1,\ast}_t S_t\Big)^+ dt \\
&\ \ \ -r_{t}^{b}\Big(Y^{h,l,x_1}_{t}B_{t}^{l}-  Z^{h,l,x_1,\ast}_t  S_t\Big)^- dt -r_{t}^{l}B_{t}^{l}Y^{h,l,x_1}_{t}\, dt + r_{t}^{l}B_{t}^{l}(Y^{h,l,x_1}_{t} - x_1)\, dt+ dA_t+dF^{C}_{t}\\
&=\widetilde{Z}^{h,l,x_1,\ast}_t \, d \wt S^{l,{\textrm{cld}}}_t
+r_{t}^{l}\Big(P^{h}_{t}+C_{t}+x_1B_{t}^{l}-(B_{t}^{l})^{-1}\widetilde{Z}^{h,l,x_1,\ast}_t S_t\Big)^+ dt \\
&\ \ \ -r_{t}^{b}\Big(P^{h}_{t}+C_{t}+x_1B_{t}^{l}-  (B_{t}^{l})^{-1}\widetilde{Z}^{h,l,x_1,\ast}_t  S_t\Big)^- dt \\
&\ \ \ -x_1 r_{t}^{l} B_{t}^{l}\, dt+r_{t}^{l}(B_t^{l})^{-1}\widetilde{Z}^{h,l,x_1,\ast}_tS_t\, dt+ dA_t-r_{t}^{c}C_{t}\, dt .
\end{align*}
Moreover, by comparing Proposition 5.2 in \cite{BR-2014} with equation (\ref{Bergman model lending BSDE}), we deduce easily that
$Y^{h,l,x_{1}}=(B^{l})^{-1} V^p(x_1,\phi , \pA ,\pC)$ and thus
\[
P^{h}=  \Blr (Y^{h,l,x_{1}} - x_{1})-C=V^p(x_1,\phi , \pA ,\pC)-x_1\Blr-C=V(x_1,\phi , \pA ,\pC)-x_1\Blr.
\]
By applying similar arguments to the counterparty's pricing problem, we obtain the equality $Y^{c,l,x_{2}}=(B^{l})^{-1}V^p(x_2,\widetilde{\phi} , -\pA ,-\pC)$, which in turn yields
\[
P^{c}(x_{2},-A,-C) =  -\Blr (Y^{c,l,x_{2}} - x_{2})+C=-V(x_2,\widetilde{\phi} , -\pA ,-\pC)+x_2\Blr.
\]
We conclude that if $C$ is given by equation \eqref{collateral 1}, then we have $C_t=q(V_{t}^{0}(x_1)-V^{h}_t)=q(-P^{h}_t)$ and thus the pair $(P^{h},\widetilde{Z}^{h,l,x_1})$ is a solution to BSDE (\ref{main BSDE hedger}).
Similarly, for $\widetilde{Z}^{c,l,x_2}:=-B_{t}^{l}Z^{c,l,x_2}$, we deduce that the pair $(P^{c},\widetilde{Z}^{c,l,x_2})$ satisfies BSDE (\ref{main BSDE counterparty}).

It remains to verify that BSDEs  (\ref{main BSDE hedger}) and (\ref{main BSDE counterparty}) are indeed well-posed. One can check that $f_{l}(t,x_1,0,0)=0$ and the mapping $f_{l}$ is uniformly $m$-Lipschitz generator (for the definition of the uniformly $m$-Lipschitz generator, see \cite{NR3}). Consequently, if $A\in\mathcal{A}(\widetilde{\mathbb{P}}^l)$ then, using Theorem 3.2 in \cite{NR3}, we conclude that BSDE (\ref{main BSDE hedger}) has a unique solution $(Y^{1},Z^{1})$ such that $(Y^{1}, m^{\ast}Z^{1}) \in \wHzero  \times \wHzerd $. Similarly, we note that
\[
g_{l}(t,x_2,0,0)=x_2 B_{t}^{l}r_{t}^{l}-r_{t}^{c}q(-Y^1_{t})-r_{t}^{l}\big(-q(-Y^1_{t})+x_2B_{t}^{l}\big)^+
+ r_{t}^{b}\big(-q(-Y^1_{t})+x_2B_{t}^{l}\big)^-
\]
where $q$ is a uniformly Lipschitz continuous function and $Y^{1}\in \wHzero  $, so that $g_{l}(t,x_2,0,0)\in \wHzero $. Moreover,
the mapping $g_{l}$ is also a uniformly $m$-Lipschitz generator, and thus BSDE (\ref{main BSDE counterparty}) has also a unique solution $(Y^{2},Z^{2})$ such that $(Y^{2}, m^{\ast}Z^{2}) \in \wHzero  \times \wHzerd $.
\end{proof}

We are now in a position to examine the range of fair bilateral prices at time $t$ (see Definition \ref{range of fair}).
It appears that, under mild assumptions, this range is non-empty when the initial endowments of the two parties have
the same sign. Let us note that this range may be empty, in general, if the initial endowments are of opposite signs, that is, when $x_{1}>0$ and $x_{2}<0$ (see Proposition \ref{Bergman model inequality proposition for positive negative initial wealth}(ii) in Section \ref{sec3.3}).

\bp \label{extended Bergman model inequality proposition for both positive initial wealth}
Let $x_{1}\ge0,\, x_{2}\ge0$ and Assumptions \ref{hypo1} and \ref{changed assumption for lending cumulative dividend price} be valid.
For any contract $(A,C)$ such that $A \in \Classapl $ we have, for every $t\in[0,T]$,
\be \label{range}
P^{c}_t (x_{2},-A,-C)\leq P^{h}_t (x_{1},A,C),  \quad \PT^l-\aass ,
\ee
so that the range of fair bilateral prices ${\cal R}^f_t(x_1,x_2)$ is non-empty, $\PT^l-\aass$
\ep

\begin{proof}
In view of Proposition \ref{hedger collateral dependent Bergman model ex-dividend price} and a suitable version of the comparison theorem for BSDEs (see Theorem 3.3 in \cite{NR3}), to establish the inequality $P^{c}_t (x_{2},-A,-C)\leq P^{h}_t (x_{1},A,C),  \ \PT^l-\aass $,
 it suffices to show that $g_l \big(t, x_{2}, Y^{1},Z^{1} \big) \ge f_l \big(t,x_{1}, Y^{1},Z^{1} \big)$, $\PT^l\otimes \Leb-\aaee$. To demonstrate the latter inequality, we denote
\bde
\delta :=g_l \big(t, x_{2}, Y^{1},Z^{1} \big)-f_l \big(t, x_{1}, Y^{1},Z^{1} \big) \\
=\rll_tB_{t}^{l} (x_{1}+x_{2})-\rll_t (\delta_{1}^{+}+\delta_{2}^{+})+\rbb_t (\delta_{1}^{-}+\delta_{2}^{-})
\ede
where
\begin{align*}
&\delta_{1}:=-Y^{1}_t-q(-Y^{1}_t)+\Blr_t x_{2}+(\Blr_t)^{-1}Z^{1,\ast}_t S_t ,
&\delta_{2}:=Y^{1}_t+q(-Y^{1}_t)+\Blr_t x_{1}-(\Blr_t)^{-1}Z^{1,\ast}_t S_t.
\end{align*}
Since, by Assumption \ref{assumption for primary assets}, the inequality $r^{l}\leq r^b$ holds, we obtain
\bde
\delta\ge\rll_tB_{t}^{l} (x_{1}+x_{2})-\rll_t (\delta_{1}+\delta_{2})=0 ,
\ede
which is the required condition.
\end{proof}

\brem
It is clear that analogous results can be established when the collateral depends only on the counterparty's value $V^{c}:=V(x_{2},\widetilde{\varphi},-A,-C)$, specifically, when Assumption \ref{hypo1} is replaced by the
postulate that $C_t=q(V^{c}_t - V_{t}^{0}(x_2))$ for some uniformly Lipschitz continuous function $q$ such that $q(0)=0$.
\erem

\brem
One can also prove similar results when the initial endowments satisfy $x_1\leq0$ and $x_2\leq0$, so that they are still of the same sign. The case where the initial endowments have opposite signs is more challenging and it is analyzed in Section  \ref{sec3.3}.
\erem

\subsubsection{European Claims in a Diffusion Model}  \label{sec3.2}

The pricing and hedging of collateralized European claims  in a diffusion model was  recently studied by Mercurio \cite{M-2013} (see also Piterbarg \cite{P10}). It should be pointed out that the hedger's and counterparty's initial endowments were implicitly assumed to be null in \cite{M-2013}. More importantly, the collateral amount for the hedger (resp., for the counterparty) was specified as a constant proportion of the hedger's (resp., the counterparty's) value, that is, it was postulated in \cite{M-2013} that $C^h=\alpha V^h$ and $C^{c}=\alpha V^{c}$ for some $\alpha\in[0,1]$. Such a specification of the margin account apparently corresponds to the situation where the hedger and the counterparty post/receive collateral of possibly different amounts to/from the third party independently of each other. Obviously, this is inconsistent with the real-life situation where the margin account is common for both parties, so that the collateral amount posted (resp., received) by one party is received (resp., posted) by another party.

For simplicity, let us assume $d=1$, so that there is only one risky asset, $S=S^1$.  This restriction can be relaxed
and thus Corollary \ref{coreur} can be easily extended to the multi-asset framework.

\bhyp \label{assdif}
 We assume that: \hfill \break
(i) the risky asset $S$ has the ex-dividend price dynamics under $\P$ given by the following expression, for $t \in [0,T]$,
\be \label{Partial netting model stock price}
dS_t = \mu(t,S_{t})\, dt +\sigma(t,S_{t})\, dW_t , \quad S_0=\ssx \in \mathcal{O},
\ee
where $W$ is a one-dimensional Brownian motion and $\mathcal{O}$ is the domain of real values that are attainable by the diffusion process $S$ (usually $\mathcal{O}=\mathbb{R}_{+}$), \hfill \break
(ii) the filtration $\gg$ is generated by the Brownian motion $W$, \hfill \break
(iii) the  coefficients $\mu$ and $\sigma$ are such that SDE (\ref{Partial netting model stock price}) has a unique strong solution, \hfill \break
(iv) the dividend process equals $\pA^1_t = \int_0^t \kappa( u, S_u) \, du $.
\ehyp

We observe that
\bde
d\wt S^{l,\textrm{cld}}_t =(\Blr_t)^{-1}\left(dS_t - \rll_t S_t \, dt + d\pA^1_t\right) =
\big( \mu(t, S_{t})+\kappa(t, S_{t})-\rll_t S_{t}\big)\, dt + \sigma (t, S_{t})\, dW_t .
\ede
We denote
\be \label{defaa}
a_{t}:=(\sigma(t, S_{t}))^{-1}\big( \mu(t, S_{t})+\kappa(t, S_{t})-\rll_t S_{t}\big)
\ee
and we suppose that $a$ satisfies Novikov's condition
\be\label{Novikov condition}
\mathbb{E}_{\mathbb{P}} \left( \exp \bigg\{ \frac{1}{2}\int_0^T|a_t|^2\, dt \bigg\} \right) <\infty.
\ee
Let us define the probability measure $\PT^l$ by setting
\be \label{defptl}
\frac{d\PT^l}{d\P}=\exp\left\{-\int_{0}^{T}a_{t}\, dW_{t}
-\frac{1}{2}\int_{0}^{T}|a_{t}|^{2}\, dt \right\}.
\ee
From the Girsanov theorem, the probability measure  $\PT^l $ is equivalent to $\P$ and the process $\wtWl$ is the Brownian motion under $\PT^l$, where $d\wtWl_{t}:=dW_{t}+a_{t}\, dt$. It is easy to see that the dynamics of the process $\wt S^{l,\textrm{cld}}$ under
$\PT^l$ are
\bde
d\wt S^{l,\textrm{cld}}_{t}=\sigma (t, S_{t})\, d\wtWl_t
\ede
and thus $\wt S^{l,\textrm{cld}}$ is a $(\PT^l , \gg)$-(local) martingale with the quadratic variation $\langle \wt S^{l,\textrm{cld}}\rangle_t=\int_{0}^t |\sigma(u, S_{u})|^{2}\, du$. Therefore, if $(\sigma(\cdot, S))^{-1}S$ is bounded, Assumption \ref{changed assumption for lending cumulative dividend price} holds, since the Brownian motion $\wtWl$ is known to have the predictable representation property under $(\gg,\PT^l)$.

Let us comment on the valuation and hedging of a European contingent claim with the hedger's collateral given by \eqref{collateral 1}.
A generic European claim pays at its expiration date $T$ the amount $H_T$ to the hedger, so that
\be \label{euro}
A_t - A_0 = H_T \I_{[T,T]}(t).
\ee
We find it convenient to denote such a contract as $(H_T,C)$.
From Proposition \ref{extended Bergman model inequality proposition for both positive initial wealth}, we deduce the following corollary.

\bcor \label{coreur}
Consider a collateralized European claim $(H_T,C)$ where the random variable $H_T$ is square-integrable under $\PT^l$.
If $x_1 \ge 0,\, x_2 \ge 0$ and Assumption \ref{hypo1} is valid, then the range of fair bilateral prices ${\cal R}^f_t(x_1,x_2)$ for $(H_T,C)$ is non-empty, $\PT^l$-a.s.
\ecor

\subsubsection{Model with an Uncertain Money Market Rate} \label{sec3.4}

 We continue working under the assumption that $x_1 \ge 0 $ and $x_2\ge0$. Let us select any $\gg$-adapted interest rate process satisfying the following condition
\be \label{rere}
\rir_t\in[r^{l}_{t},r^{b}_t] \ \text{ for every } t\in[0,T].
\ee
We preserve all other assumptions regarding the market model at hand, including the set of traded risky assets but, for the sake of
comparison, we now also consider an additional market model with the single uncertain money market rate $\rir $. To be more specific,  part (iv) in Assumption \ref{assumption for primary assets} becomes: $dB^{l}_{t}=\rir_tB^{l}_{t}\, dt$, $dB^{b}_{t}= \rir_tB^{b}_{t}\, dt$ and $dB^{c}_{t}=r^{c}_{t}B^{c}_{t}\, dt$  for some $\mathbb{G}$-adapted and bounded processes $\rir ,\, r^{l}$ and $r^c$.
Under these assumptions,  the hedger and the counterparty have the same ex-dividend price  $\Prir$, which does not depend on their initial endowments. Intuitively, this is due to the fact that the situation in now fully symmetric since we deal now with a
single interest rate. Formally,  the ex-dividend price process $\Prir=Y$ now coincides with the unique solution to the BSDE
\be \label{main BSDE linear}
\left\{ \begin{array}
[c]{l}
dY_t = Z^{\ast}_t \, d \wt S^{l,{\textrm{cld}}}_t +f(t, Y_t, Z_t)\, dt + dA_t, \medskip\\
Y_T=0 ,
\end{array} \right.
\ee
where the generator $f$ is given by the following expression
\bde
f(t,y,z)= (r_{t}^l - \rir_t) (B_{t}^{l})^{-1}z^\ast S_t -r_{t}^{c}q(-y)+\rir_t(y+q(-y)).
\ede

The next result is not only more general but, in our opinion, it is also more natural than Proposition 4.1 in Mercurio~\cite{M-2013} where, as was already mentioned at the beginning of Section \ref{sec3.2}, the collateral amount for each party was tied to his unilateral value of the contract. Note that the prices $P^{h}(0,A,C)$ and $P^{c}(0,-A,-C)$ are computed in Bergman's model with differential borrowing and
lending rates $\rll $ and $\rbb$ under the assumption that $x_1=x_2=0$ and the collateral $C$ is given by \eqref{collateral 1}.

\bp \label{link to economies with no differential rates}
Consider the market model with the money market rate $r$.  \hfill \break
 (i) The price $\Prir$ of any contract $(A,C)$ such that $A \in \Classapl $ satisfies
 $\Prir\leq P^{h}(0,A,C),$ $ \PT^l-\aass$ \hfill \break
 (ii) If $x_1=x_2=0$ and the function $q$ in equation \eqref{collateral 1} satisfies $(\rir_t-r_{t}^{c})(q(y_1)-q(y_2))\leq 0$ for all $y_1\ge y_2$, then also $P^{c}(0,-A,-C) \leq \Prir,$ $\PT^l-\aass$
\ep

\begin{proof}
(i) In view of the comparison theorem for BSDEs (see Theorem 3.3 in \cite{NR3}), it is sufficient to show that the
inequality $f_{l}\big(t, 0, Y_t,Z_t \big) \leq f \big(t, Y_t,Z_t \big)\leq g_{l}\big(t, 0, Y_t,Z_t \big)$ holds $\PT^l\otimes \Leb-\aaee$. Let us denote
\bde
\delta_{1}:=Y_t+q(-Y_t)+\Blr_t x_{1}-(B_{t}^{l})^{-1}Z^{\ast}_t S_t.
\ede
From the assumption that $\rir_t\in[r^{l}_{t},r^{b}_t]$ for all $t \in [0,T]$, we obtain
\begin{align*}
\delta &:=f_l\big(t, x_{1}, Y_t,Z_t \big)-f \big(t, Y_t,Z_t \big) =\rir_t(B_{t}^{l})^{-1}Z^\ast_t S_t -x_1 B_{t}^{l}r_{t}^{l}+r_{t}^{l}\delta_1^+-r_{t}^{b}\delta_1^--\rir_tY_t-\rir_tq(-Y_t)\\
&\leq \rir_t(B_{t}^{l})^{-1}Z^\ast_t S_t -x_1 B_{t}^{l}r_{t}^{l}+\rir_t\delta_1-\rir_t(Y_t+q(-Y_t))=(\rir_t-r_{t}^{l})x_{1}B_{t}^{l}.
\end{align*}
Therefore, if  $x_1=0$, then $\delta\leq0$. Consequently, $Y\leq Y^{1}$ and thus $ \Prir\leq P^{h}(0,A,C)$. \vskip 3 pt
\noindent (ii) We now assume that the hedger's initial endowment is null and we examine the pricing problem for the counterparty.
Recall that we postulate that $C_t=q(-Y^1)=q(-P^{h}(0,A,C))$. Let us denote
\bde
\delta_{2}:=-Y_t-q(-Y_t^1)+\Blr_t x_{2}+(B_{t}^{l})^{-1}Z^{\ast}_t S_t .
\ede
From $\rir_t\in[r^{l}_{t},r^{b}_t]$, we obtain
\begin{align*}
&\widetilde{\delta} :=f \big(t, Y_t,Z_t \big)-g_{l}\big(t, x_{2}, Y_t,Z_t \big)\\
&=-\rir_t(B_{t}^{l})^{-1}Z^\ast_t S_t -r_{t}^{c}q(-Y_t)+r_{t}^{c}q(-Y_t^1)+\rir_t(Y_t+q(-Y_t))-x_2 B_{t}^{l}r_{t}^{l}+r_{t}^{l}\delta_2^+-r_{t}^{b}\delta_2^-\\
&\leq -\rir_t(B_{t}^{l})^{-1}Z^\ast_t S_t  -r_{t}^{c}q(-Y_t)+r_{t}^{c}q(-Y_t^1)+\rir_t(Y_t+q(-Y_t))-x_2 B_{t}^{l}r_{t}^{l}+\rir_t\delta_2\\
&=(\rir_t-r_{t}^{l})x_{2}B_{t}^{l}+(\rir_t-r_{t}^{c})(q(-Y_t)-q(-Y_t^1)).
\end{align*}
Since $(\rir_t-r_{t}^{c})(q(y_1)-q(y_2))\leq 0$ for all $y_1\ge y_2$ and $Y\leq Y^{1}$, we have $(\rir_t-r_{t}^{c})(q(-Y_t)-q(-Y_t^1))\leq0$. Therefore, if $x_2=0$, then $\widetilde{\delta}\leq0$.
We conclude that $P^{c}(0,-A,-C) \leq \Prir $.
\end{proof}

\brem
Note that the condition $(\rir_t-r_{t}^{c})(q(y_1)-q(y_2))\leq 0$ for all $y_1\ge y_2$ can be easily ensured. For instance, if $q$ is an increasing function (e.g., the one given in Example \ref{ex1} when $\alpha_1, \alpha_2 > -1 $), then it suffices to postulate
that $\rir_t\leq r_{t}^{c}$.  When $x_1 \ge 0$ and $x_2\ge0$, it is not clear whether the inequalities $P^{c}(x_2,-A,-C) \leq \Prir\leq P^{h}(x_1,A,C)$ are valid. Indeed, Proposition  \ref{link to economies with no differential rates} only shows that they are valid under the assumption that $x_1=x_2=0$.
\erem

\brem
As in Section 5.4 of \cite{NR2}, one can prove the monotonicity and stability properties of the price with respect to the initial endowment of each party. No difficulty arises in the case of the hedger's price. Since the counterparty's price $P^c(x_1,x_2,-A,-C)$ depends also on the hedger's initial endowment, the arguments used in \cite{NR2} should be slightly modified.
\erem

\subsection{Initial Endowments of Opposite Signs} \label{sec3.3}

So far, we worked under the assumption that the initial endowments of both parties are non-negative.
We will now briefly examine the situation where $x_{1}\ge0$ and $x_{2}\leq 0$. As in our previous work \cite{NR2},
the concept of a `martingale measure' will now be specified in a more abstract way than in Assumption
\ref{assumption for lending cumulative dividend price} by making reference to some auxiliary process, which is hereafter
denoted as $\beta $.

\bhyp \label{changed assumption for artifical cumulative dividend price}
We postulate that: \hfill \break
(i) there exists a probability measure $\PTb $ equivalent to $\P$ such that the processes $\wt S^{i,\textrm{cld}},\, i=1,2, \dots ,d$, which are given by
\be \label{auxiliary processes}
d\wt S^{i,\textrm{cld}}_t = dS^i_t + d\pA^i_t - \beta^{i}_{t}S_{t}^{i}\,dt
\ee
for some $\rr^d$-valued, $\gg$-adapted, bounded processes $\beta$ satisfying $r^{b}\leq\beta^{i}$ for $i=1,\dots ,d$, are $(\PTb , \gg)$-continuous, square-integrable martingales and have the predictable representation property with respect to the filtration $\gg$ under~$\PTb$, \hfill \break
(ii) there exists an $\mathbb{R}^{d\times d}$-valued, $\gg$-adapted process $m$ such that
\be \label{auxiliary processes quadratic variation}
\langle \wt S^{\textrm{cld}}\rangle_{t}=\int_{0}^{t}m_{u}m_{u}^{\ast}\,du
\ee
where $m m^{\ast}$ is invertible and satisfies $m m^{\ast}= \mathbb{S}\sigma\sigma^{\ast}\mathbb{S}$.  Here $\sigma$ is a $d$-dimensional square matrix of $\gg$-adapted processes, which satisfies the ellipticity condition \eqref{elli}.
\ehyp

The following proposition establishes the no-arbitrage property of Bergman's model under the present assumptions.
Since the proof of this result is very similar to the proof of Proposition 3.2 in \cite{NR2}, it is omitted.

\bp \label{remark for non-arbitrage model}
If Assumption \ref{changed assumption for artifical cumulative dividend price} holds, then
Bergman's model is arbitrage-free for the hedger and the counterparty in respect of any initial endowments and any contract $(A,C)$.
\ep

For the sake of comparison, we recall Proposition 3.3 from \cite{NR4} concerning the case of exogenous collateral.

\bp \label{Bergman model general pricing proposition}
Let $x_{1}\ge0,\, x_{2}\leq0$ and Assumption \ref{changed assumption for artifical cumulative dividend price} be valid.
Then for any contract $(A,C)$ such that $A^C \in \Classaptb $ we have $P^{h}(x_{1},A,C)=\widetilde{Y}^{h,x_{1}}-C$ and $P^{c}(x_{2},-A,-C)=\widetilde{Y}^{c,x_{2}}-C$ where  $(\widetilde{Y}^{h,x_{1}},\widetilde{Z}^{h,x_{1}})$ is the unique
solution to the following BSDE
\bde
\left\{ \begin{array}
[c]{ll}
d\widetilde{Y}^{h,x_{1}}_t =\widetilde{Z}^{h,x_{1},\ast}_t\, d\wt S_t^{\textrm{cld}}+G^{h}(t,x_{1}, \widetilde{Y}^{h,x_{1}}_t,\widetilde{Z}^{h,x_{1}}_t)\,dt + dA^C_t, \medskip \\
\widetilde{Y}^{h,x_{1}}_T=0,
\end{array} \right.
\ede
and $(\widetilde{Y}^{c,x_{2}},\widetilde{Z}^{c,x_{2}})$ is the unique solution to the following BSDE
\bde
\left\{ \begin{array}
[c]{ll}
d\widetilde{Y}^{c,x_{2}}_t=\widetilde{Z}^{c,x_{2},\ast}_t\, d\wt S_t^{\textrm{cld}}+G^{c}(t, x_{2}, \widetilde{Y}^{c,x_{2}}_t,\widetilde{Z}^{c,x_{2}}_t)\,dt + dA^C_t, \medskip \\
\widetilde{Y}^{c,x_{2}}_T=0,
\end{array} \right.
\ede
where the generators $G^{h}$ and $G^{c}$ are given by the following expressions
\bde
G^{h}(t,x_1,y,z):=\sum_{i=1}^dz^{i}\beta^{i}_{t}S_{t}^{i}-x_1\rll_t\Blr_t+\rll_t \Big(y+x_1\Blr_t- z^{\ast}S_t\Big)^+- \rbb_t \Big( y+x_1\Blr_t-z^{\ast}S_t \Big)^-
\ede
and
\bde
G^{c}(t,x_2,y,z):=\sum_{i=1}^dz^{i}\beta^{i}_{t}S_{t}^{i}+x_2\rbb_t\Bbr_t-\rll_t \Big(-y+x_2\Bbr_t+z^{\ast}S_t\Big)^++ \rbb_t \Big( -y+x_2\Bbr_t+z^{\ast}S_t \Big)^-.
\ede
\ep

The next result covers the case of hedger's collateral when the initial endowments $x_1$ and $x_2$ have opposite signs. Since its proof is analogous to that of Proposition \ref{hedger collateral dependent Bergman model ex-dividend price}, it is not presented here.

\bp \label{opposite sign hedger collateral dependent Bergman model ex-dividend price}
Let $x_{1}\ge0, \, x_{2}\leq0$ and Assumptions \ref{hypo1} and \ref{changed assumption for artifical cumulative dividend price} be valid.
For any contract $(A,C)$ such that $A \in \Classaptb $, the hedger's ex-dividend price equals $P^h= \overline{Y}^1$ where $(\overline{Y}^{1}, \overline{Z}^{1})$ is the unique solution to the BSDE
\be \label{main BSDE hedger 1}
\left\{ \begin{array}
[c]{l}
d\overline{Y}_t^1 = \overline{Z}^{1,\ast}_t \, d \wt S^{{\textrm{cld}}}_t
+\overline{f} \big(t, x_1,\overline{Y}^1_t, \overline{Z}^1_t \big)\, dt + dA_t, \medskip\\
\overline{Y}^1_T=0,
\end{array} \right.
\ee
with the generator $\overline{f}$ given by
\begin{align*}
\overline{f}(t,x_1,y,z)=&\sum_{i=1}^dz^{i}\beta^{i}_{t}S_{t}^{i}-x_1\rll_t\Blr_t-r_{t}^{c}q(-y)\\
&\mbox{}+r_{t}^{l}\Big(y+q(-y)+x_1B_{t}^{l}-z^\ast S_t\Big)^+-r_{t}^{b}\Big(y+q(-y)+x_1B_{t}^{l}-z^\ast S_t\Big)^-
\end{align*}
 and the counterparty's ex-dividend price equals $P^c= \overline{Y}^2$ where $(\overline{Y}^{2}, \overline{Z}^{2})$ is the unique solution to the BSDE
\be \label{main BSDE counterparty 1}
\left\{ \begin{array}
[c]{l}
d\overline{Y}_t^2 = \overline{Z}^{2,\ast}_t \, d \wt S^{{\textrm{cld}}}_t
+\overline{g} \big(t, x_2,\overline{Y}^2_t, \overline{Z}^2_t \big)\, dt + dA_t, \medskip\\
\overline{Y}^2_T=0,
\end{array} \right.
\ee
with the generator $\overline{g}$ given by
\begin{align*}
\overline{g}(t,x_2,y,z)=&\sum_{i=1}^dz^{i}\beta^{i}_{t}S_{t}^{i}+x_2\rbb_t\Bbr_t-r_{t}^{c}q(-\overline{Y}^1_{t})\\
&\mbox{}\ -r_{t}^{l}\Big(-y-q(-\overline{Y}^1_{t})+x_2B_{t}^{b}+z^\ast S_t\Big)^++r_{t}^{b}\Big(-y-q(-\overline{Y}^1_{t})+x_2B_{t}^{b}+z^\ast S_t\Big)^-.
\end{align*}
\ep

We are now in a position to analyze the range of fair bilateral prices when the initial endowments of
counterparties are of opposite signs.

\bp \label{Bergman model inequality proposition for positive negative initial wealth}
Let $x_{1}\ge0,\, x_{2}\leq0$ and Assumptions \ref{hypo1} and \ref{changed assumption for artifical cumulative dividend price} be valid.  \hfill \break
(i) If $x_{1}x_{2}=0$, then for any contract $(A,C)$ such that $A \in \Classaptb $ we have, for all $t\in[0,T]$,
\be \label{equ1}
P^{c}_t (x_{2},-A,-C)\leq P^{h}_t (x_{1},A,C),  \quad \PTb-\aass
\ee
(ii) Let $r^{l}$ and $r^{b}$ be deterministic and satisfy $r^{l}_{t}<r^{b}_{t}$ for all $t\in[0,T]$. Then
\eqref{equ1} holds for all contracts $(A,C)$ such that $A \in \Classaptb $ and all $t\in[0,T]$
if and only if $x_{1}x_{2}=0$.
\ep

\begin{proof}
(i) We consider solutions $(\overline{Y}^1,\overline{Z}^1)$ and  $(\overline{Y}^2,\overline{Z}^2)$ to BSDEs
\eqref{main BSDE hedger 1} and \eqref{main BSDE counterparty 1} studied in Proposition \ref{opposite sign hedger collateral dependent Bergman model ex-dividend price} and we wish to apply the comparison theorem for BSDEs to
show that $\overline{Y}^1 \geq \overline{Y}^2$.    We claim that if $x_{1}\ge0$ and $x_{2}\leq0$, then
\be
\delta:=\overline{g}(t,x_{2},\overline{Y}^{1}_t,\overline{Z}^{1}_t)-\overline{f}(t,x_{1},\overline{Y}^{1}_t,\overline{Z}^{1}_t) \ge \max \big\{-(\rbb_t- \rll_t)x_{1}B_{t}^{l},\, (\rbb_t- \rll_t )x_{2}B_{t}^{b} \big\}. \label{rre}
\ee
Indeed, we have
\bde
\delta =x_{1}r_{t}^{l}B_{t}^{l}+x_{2}r_{t}^{b}B_{t}^{b} - \rll_t (\delta_{1}^{+}+\delta_{2}^{+}) +
\rbb_t (\delta_{1}^{-}+\delta_{2}^{-})
\ede
where we denote
\begin{align*}
&\delta_{1}:=-\overline{Y}_{t}^{1}-q(-\overline{Y}_{t}^{1})+x_2 B_{t}^{b}+\overline{Z}^{1,\ast}_t S_t , \\
&\delta_{2}:=\overline{Y}_{t}^{1}+q(-\overline{Y}_{t}^{1})+x_1 B_{t}^{l}-\overline{Z}^{1,\ast}_t S_t.
\end{align*}
From the postulated inequality $\rll_{t}\leq\rbb_t$, it follows easily that
\bde
\delta\ge x_{1}r_{t}^{l}B_{t}^{l}+x_{2}r_{t}^{b}B_{t}^{b} - \rll_t (\delta_{1}+\delta_{2}) =(\rbb_t- \rll_t )x_{2}B_{t}^{b}
\ede
and
\bde
\delta\ge x_{1}r_{t}^{l}B_{t}^{l}+x_{2}r_{t}^{b}B_{t}^{b} - \rbb_t (\delta_{1}+\delta_{2})= -(\rbb_t- \rll_t)x_{1}B_{t}^{l}.
\ede
We have thus proven that \eqref{rre} is valid.
If $x_{1}x_{2}=0$, then the right-hand side in \eqref{rre} is non-negative. Hence $\delta\ge0$ and thus, from the comparison theorem for BSDEs and the equality $(P^h, P^c) = (\overline{Y}^{1}, \overline{Y}^{2})$   (see Proposition \ref{opposite sign hedger collateral dependent Bergman model ex-dividend price}), we deduce that \eqref{equ1} is satisfied for every $t\in[0,T]$. \vskip 5 pt
\noindent (ii) We now assume that the interest rates $r^{l}$ and $r^{b}$ are deterministic and satisfy $r^{l}_{t}<r^{b}_{t}$ for all $t\in[0,T]$. If $x_{1}x_{2} \neq 0$, then the example from the proof of Proposition 5.4 in \cite{NR2} gives a contract $(A,C)$ with $q\equiv0$, such that the inequality $P^{c}_{0} (x_{2},-A,-C) >  P^{h}_{0} (x_{1},A,C)$, $\PTb-\aass$,
holds in the present framework, so that the set ${\cal R}^f_0(x_1,x_2)$ is empty.
\end{proof}

%

\section{Bergman's Model with Negotiated Collateral} \label{sec4}

Our objective in Sections \ref{sec4} and \ref{sec6} is to analyze the situation where the collateral amount $C$ relies upon both the hedger's value $V^{h}:=V(x_{1},\varphi,A,C)$ and the counterparty's value $V^{c}:=V(x_{2},\widetilde{\varphi},-A,-C)$. Specifically,
Assumption \ref{hypo1} is replaced by the following postulate in which the collateral amount may depend on the contract's valuation by both parties. For convenience, we then say that the collateral is {\it negotiated} by the two parties, the sense that both the choice of
the collateral convention  $\Cnq $ and the dynamic computation of the collateral amount $C_t$ involve both parties of a contract, in general.
\bhyp \label{hypo2}
{\rm  The {\it negotiated collateral} $C$ is given by
\be \label{collateral 2}
C_t=\Cnq \big( V_{t}^{0}(x_1)-V^{h}_t, V^{c}_t - V_{t}^{0}(x_2) \big)
\ee
where $\Cnq:\mathbb{R}^{2}\rightarrow\mathbb{R}$ is a uniformly Lipschitz continuous function such that $\Cnq(0,0)=0$.}
\ehyp

The case of a negotiated collateral should be contrasted with the situation considered in the preceding section,
where it was postulated that the collateral amount was set by one party only. Let us observe that the prices for both
parties will now depend on the vector of initial endowments $(x_1,x_2)$, but we will keep writing $P^{h}(x_{1},A,C)$ and $P^{c}(x_2,-A,-C)$ instead of $P^{h}(x_{1},x_2,A,C)$ and $P^{c}(x_{1},x_2,-A,-C)$, respectively.
For $x_1\ge0$ and $x_2\ge0$, using the arguments from the proof of Proposition \ref{hedger collateral dependent Bergman model ex-dividend price}, we obtain
\bde
P^{h}=P^{h}(x_1,A,C) =V(x_1,\phi , \pA ,\pC)-x_1\Blr=V^{h}-x_1\Blr
\ede
and
\bde
P^{c}=P^{c}(x_2,-A,-C) = -V(x_2,\widetilde{\phi} , -\pA ,-\pC)+x_2\Blr=-V^{c}+x_2\Blr.
\ede
Similarly, for $x_2\leq 0$, we have
\bde
P^{c}=P^{c}(x_2,-A,-C) = -V(x_2,\widetilde{\phi} , -\pA ,-\pC)+x_2\Bbr=-V^{c}+x_2\Bbr.
\ede
We conclude that the following equality is valid, for $x_1 \geq 0 $ and an arbitrary $x_2$,
\be
C_t=\Cnq \big( V_{t}^{0}(x_1)-V^{h}_t,V^{c}_t- V_{t}^{0}(x_2)\big)=\Cnq (-P^{h}_t,-P^{c}_t).
\ee

\bex \label{ex2}
As a particular instance of equation \eqref{collateral 2}, we can consider the convex collateralization given by $\Cnq (y_1,y_2)=\alpha y_1+(1-\alpha)y_2$ for some $\alpha\in[0,1]$, so that
\bde
C_{t}=\alpha(V_{t}^{0}(x)-V_{t}^h)+(1-\alpha)(V_{t}^c - V_{t}^{0}(x)) = - (\alpha P^h_t + (1 -\alpha )P^c_t ).
\ede
\eex

\subsection{Fully-Coupled Pricing BSDE} \label{sec4.1}

The following result, which covers the case of non-negative initial endowments, is a rather straightforward extension of Proposition \ref{hedger collateral dependent Bergman model ex-dividend price} and thus its proof is omitted. It is worth noting that the processes $Y$
and $Z$ are $\rr^2$-valued and $\rr^{d \times 2}$-valued, respectively.

\bp \label{symmetric collateral dependent Bergman model ex-dividend price}
Let $x_1\ge0,\, x_2\ge0$ and Assumptions  \ref{changed assumption for lending cumulative dividend price} and \ref{hypo2} be valid.
For any contract $(A,C)$ such that $A \in \Classapl $, the hedger's and counterparty's ex-dividend prices satisfy $(P^{h},P^{c})^\ast=Y$ where the pair $(Y,Z)$ solves the following two-dimensional, fully-coupled BSDE
\be \label{main two dimensioanl BSDE}
\left\{ \begin{array}
[c]{l}
dY_t = Z^{\ast}_t \, d \wt S^{l,{\textrm{cld}}}_t
+ g \big(t,Y_t, Z_t \big)\, dt + d\overline{A}_t, \medskip\\
Y_T=0 ,
\end{array} \right.
\ee
where $g=(g^1,g^2)^\ast$, $\overline{A}=(A,A)^\ast$ and, for all $y=(y_1,y_2)^\ast\in\mathbb{R}^{2},\, z=(z_1,z_2)\in\mathbb{R}^{d\times 2}$,
\begin{align} \label{generator 1}
g^{1}(t,y,z)&=r_{t}^{l}(B_t^{l})^{-1}z^{\ast}_1 S_t -x_1 B_{t}^{l}r_{t}^{l}-r_{t}^{c}\Cnq (-y_1,-y_2)\nonumber \\
&\mbox{}\ \ +r_{t}^{l}\Big(y_1+\Cnq (-y_1,-y_2)+x_1B_{t}^{l}-(B_{t}^{l})^{-1}z^{\ast}_1 S_t\Big)^+ \\
&\mbox{}\ \ -r_{t}^{b}\Big(y_1+\Cnq (-y_1,-y_2)+x_1B_{t}^{l}-(B_{t}^{l})^{-1}z^{\ast}_1 S_t\Big)^-  \nonumber
\end{align}
and
\begin{align} \label{generator 2}
g^{2}(t,y,z)&=r_{t}^{l}(B_t^{l})^{-1}z^{\ast}_2 S_t +x_2 B_{t}^{l}r_{t}^{l}-r_{t}^{c}\Cnq (-y_1,-y_2)\nonumber \\
&\mbox{}\ \ -r_{t}^{l}\Big(-y_2-\Cnq (-y_1,-y_2)+x_2B_{t}^{l}+(B_{t}^{l})^{-1}z^{\ast}_2 S_t\Big)^+\medskip\\
&\mbox{}\ \ +r_{t}^{b}\Big(-y_2-\Cnq (-y_1,-y_2)+x_2B_{t}^{l}+(B_{t}^{l})^{-1}z^{\ast}_2 S_t\Big)^-. \nonumber
\end{align}
\ep

Obviously, the prices for both parties depend here on the vector $(x_1,x_2)$ of initial endowments, so that the notation $P^h =P^{h}(x_{1},x_2,A,C)$ and $P^c= P^{c}(x_{1},x_2,-A,-C)$ would be more appropriate. However, for brevity, they will still be denoted as $P^{h}(x_{1},A,C)$ and $P^{c}(x_2,-A,-C)$, respectively. The case of initial endowments of opposite signs is covered by the following proposition, which corresponds to Proposition \ref{opposite sign hedger collateral dependent Bergman model ex-dividend price}.

\bp \label{opposite sign symmetric collateral dependent Bergman model ex-dividend price}
Let $x_1\ge0,\, x_2\le0$ and Assumptions  \ref{changed assumption for artifical cumulative dividend price} and \ref{hypo2} be valid.  For any contract $(A,C)$ such that $A \in \Classaptb $, the hedger's and counterparty's ex-dividend prices satisfy $(P^{h},P^{c})^\ast=\widehat{Y}$ where the pair $(\widehat{Y},\widehat{Z})$ solves the following two-dimensional fully-coupled BSDE
\be \label{main two dimensioanl BSDE 2}
\left\{ \begin{array}
[c]{l}
d\widehat{Y}_t = \widehat{Z}^{\ast}_t \, d \wt S^{{\textrm{cld}}}_t
+\widehat{g}\big(t,\widehat{Y}_t,\widehat{ Z}_t \big)\, dt + d\overline{A}_t, \medskip\\
\widehat{Y}_T=0,
\end{array} \right.
\ee
where $\widehat{g}=(\widehat{g}^1,\widehat{g}^2)^\ast$, $\overline{A}=(A,A)^\ast$ and, for all $y=(y_1,y_2)^\ast\in\mathbb{R}^{2}$ and $z=(z_1,z_2)\in\mathbb{R}^{d\times 2}$,
\begin{align} \label{generator 3}
\widehat{g}^{1}(t,y,z)&=\sumik_{i=1}^dz^{i}_1\beta^{i}_{t}S_{t}^{i} -x_1 B_{t}^{l}r_{t}^{l}-r_{t}^{c}\Cnq (-y_1,-y_2)\nonumber\\
&\mbox{}\ \ +r_{t}^{l}\Big(y_1+\Cnq (-y_1,-y_2)+x_1B_{t}^{l}-z^{\ast}_1 S_t\Big)^+\\
&\mbox{}\ \ -r_{t}^{b}\Big(y_1+\Cnq (-y_1,-y_2)+x_1B_{t}^{l}-z^{\ast}_1 S_t\Big)^- \nonumber
\end{align}
and
\begin{align} \label{generator 4}
\widehat{g}^{2}(t,y,z)&=\sumik_{i=1}^dz^{i}_2\beta^{i}_{t}S_{t}^{i} +x_2 B_{t}^{b}r_{t}^{b}-r_{t}^{c}\Cnq (-y_1,-y_2)\nonumber\\
&\mbox{}\ \ -r_{t}^{l}\Big(-y_2-\Cnq (-y_1,-y_2)+x_2B_{t}^{b}+z^{\ast}_2 S_t\Big)^+\\
&\mbox{}\ \ +r_{t}^{b}\Big(-y_2-\Cnq (-y_1,-y_2)+x_2B_{t}^{b}+z^{\ast}_2 S_t\Big)^-.\nonumber
\end{align}
\ep

\begin{proof}
 Once again, the proof is similar to the proof of Proposition \ref{hedger collateral dependent Bergman model ex-dividend price}. We also use Theorem 3.2 in \cite{NR3} to show the well-posedness of BSDEs (\ref{main two dimensioanl BSDE}) and (\ref{main two dimensioanl BSDE 2}). Although the BSDE studied in \cite{NR3} is one-dimensional, it is clear that Theorem 3.2 in \cite{NR3} can be easily extended to the multi-dimensional framework.
\end{proof}

\subsection{Backward Stochastic Viability Property}  \label{sec4.2}

To obtain the range of fair bilateral prices in the case of the negotiated collateral, one needs to compare the two components of a solution to fully-coupled BSDEs (\ref{main two dimensioanl BSDE}) and (\ref{main two dimensioanl BSDE 2}). When these BSDE are driven by a general continuous martingale, this is a challenging open problem.  Fortunately, in most commonly used financial models,  the pricing BSDEs are in fact driven by a Brownian motion. Under this assumption, using the ideas from Hu and Peng \cite{HP-2006} and the characterization for the backward stochastic viability property (BSVP) given by Buckdahn et al. \cite{BQR-2000},
we will be able to compare the two one-dimensional components, $Y^1$ and $Y^2$, of a unique solution to BSDE (\ref{main two dimensioanl BSDE}) by producing first a suitable version of component-wise comparison theorem (see Theorem  \ref{comparison theorem for coupled BSDE}  below).

Let us first recall the definition of the {\it backward stochastic viability property} (BSVP, for short), which was studied
by Buckdahn et al. \cite{BQR-2000}. Let $(\Omega , {\cal F}, \P )$ be a probability space endowed with the filtration
 $\ff$ generated by a $d$-dimensional Brownian motion $W$. For any Euclidean space $\mathcal{H}$, we denote $L^{2}_{ad}(\Omega,C([0,T],\mathcal{H}))$ by the closed, linear subspace of $\ff$-adapted processes of the space $L^{2}(\Omega,\mathcal{F},\mathbb{P},C([0,T],\mathcal{H}))$.
Also, let $L^{2}_{ad}(\Omega\times(0,T),\mathcal{H})$ be the Hilbert space of $\ff$-adapted and measurable processes $X$ such that $\|X\|_{2}=\left(\mathbb{E}\int_{0}^{T}|X_{t}|^{2}\, dt\right)^{1/2}<\infty$.
We now consider the following $n$-dimensional BSDE
\be \label{Brownian BSDE}
Y_{t}=\eta+\int_{t}^{T}h(s,Y_s, Z_s)\, ds - \int_{t}^{T}Z_s \, dW_s
\ee
where $\eta$ is an $\mathbb{R}^n$-valued random variable and the generator $h$ satisfies the following assumption.

\bhyp \label{assumption for Brownian BSDE generator}
Let the mapping $h:\Omega\times[0,T]\times\mathbb{R}^{n}\times\mathbb{R}^{n \times d}\rightarrow\mathbb{R}^{n}$ satisfy: \hfill \break
(i)  $\mathbb{P}$-a.s., for all $(y,z)\in\mathbb{R}^{n}\times\mathbb{R}^{n \times d}$, the process $(h(t,y,z))_{t\in[0,T]}$ is $\ff$-adapted and the mapping $t \to h(t,y,z)$ is continuous, \hfill \break
(ii) the function $h$ is uniformly Lipschitzian with respect to $(y,z)$: there exists a constant $L\ge0$ such that $\mathbb{P}$-a.s. for all $t\in[0,T]$ and $y,y^{\prime}\in\mathbb{R}^n,\, z,z^{\prime}\in\mathbb{R}^{n \times d}$
\[
|h(t,y,z)-h(t,y^{\prime},z^{\prime})|\leq L(|y-y^{\prime}|+|z-z^{\prime}|),
\]
(iii) the random variable $\sup_{\, t\in[0,T]}|h(t,0,0)|^{2}$ is square-integrable under $\P$.
\ehyp

The following definition is due to Buckdahn et al. \cite{BQR-2000}.

\bd
We say that BSDE (\ref{Brownian BSDE}) has the  {\it backward stochastic viability property} (BSVP) in $K$ if and only if: for any $U\in [0,T]$ and an arbitrary $\eta\in L^{2}(\Omega,\mathcal{F}_U,\mathbb{P};K)$, the unique solution $(Y,Z)\in L^{2}_{ad}(\Omega,C([0,U],\mathbb{R}^n))\times L^{2}_{ad}(\Omega\times(0,U),\mathbb{R}^{n \times d} )$ to the BSDE (\ref{Brownian BSDE}) over time interval $[0,U]$, that is,
\be \label{Brownian BSDE 2}
Y_{t} = \eta + \int_{t}^{U} h(s,Y_s, Z_s)\, ds- \int_{t}^{U} Z_s \, dW_s,
\ee
satisfies $Y_t \in K$ for all $t\in[0,U]$, $\mathbb{P}$-a.s.
\ed

For a non-empty, closed, convex set of $K\subset\mathbb{R}^{n}$, let $\Pi_{K}(y)$ be the projection of a point $y \in \mathbb{R}^{n}$ onto $K$ and let $d_{K}(y)$ be the distance between $y$ and $K$. The following result was established by Buckdahn et al. \cite{BQR-2000}.

\bp \label{BSVP proposition}
Let the generator $h$ satisfy Assumption \ref{assumption for Brownian BSDE generator}. Then BSDE (\ref{Brownian BSDE}) has the BSVP in $K$ if and only if for any $t\in[0,T]$, $z\in\mathbb{R}^{n \times d}$ and $y\in\mathbb{R}^n$ such that $d^{2}_{K}(\cdot)$ is twice differentiable at $y$, we have
\be \label{bsvp1}
4\langle y-\Pi_{K}(y),h(t,\Pi_{K}(y),z)\rangle\leq \langle D^{2}d_{K}^2(y)z,z\rangle +Md^{2}_{K}(y)
\ee
where $M>0$ is a constant independent of $(t,y,z)$.
\ep

Motivated by results from Hu and Peng \cite{HP-2006}, we will show that Proposition \ref{BSVP proposition} can be used to establish a convenient version of component-wise comparison theorem for the two-dimensional BSDE. Specifically, we prove the following theorem, in which
we denote $Y=(Y^1,Y^{2})^\ast ,\, Z=(Z^1,Z^{2})^\ast$ and
\bde
h(t,y,z)=\big( h^{1}(t,y^1,y^2,z^1,z^2),h^2(t,y^1,y^2,z^1,z^2) \big)^\ast .
\ede

\bt \label{comparison theorem for coupled BSDE}
Consider the two-dimensional BSDE
\be \label{two dimensioanl Brownian BSDE}
Y_{t} = \eta + \int_{t}^{T}h(s,Y_s, Z_s)\, ds - \int_{t}^{T}Z_s \, dW_s
\ee
driven by the $d$-dimensional Brownian motion $W$, where the generator $h =(h^{1},h^{2})^*$ satisfies Assumption \ref{assumption for Brownian BSDE generator}. The following statements are equivalent: \hfill \break
(i) for any $U\in[0,T]$ and $\eta^1,\eta^2\in L^{2}(\Omega,\mathcal{F}_U,\mathbb{P},\mathbb{R})$ such that $\eta^1\ge \eta^2$, the unique solution $(Y,Z)\in L^{2}_{ad}(\Omega,C([0,U],\mathbb{R}^2))\times L^{2}_{ad}(\Omega\times(0,U),\mathbb{R}^{2 \times d})$ to (\ref{two dimensioanl Brownian BSDE}) on $[0,U]$  satisfies $Y^1_t\ge Y^2_t$ for all $t \in [0,U]$, \hfill \break
(ii) the following inequality holds, for all $y_1,y_{2}\in \mathbb{R}$ and $z_1,z_{2}\in\mathbb{R}^d$,
\be \label{BSVP condition}
\begin{array}[c]{l}
-4y_1^-[h^{1}(t,y_1^++y_2,y_2,z_1+z_2,z_2)-h^{2}(t,y_1^++y_2,y_2,z_1+z_2,z_2)]\medskip\\
\leq M|y_1^-|^2+2   |z_1|^2 \I_{\{y_1<0\}}, \quad \mathbb{P}-a.s.
\end{array}
\ee
\et

\begin{proof}
Let us denote $\widetilde{Y}=(Y^1-Y^2,Y^{2})^\ast,\, \widetilde{Z}=(Z^1-Z^2,Z^{2})^\ast,\,
\widetilde{\eta}=(\eta^1-\eta^2,\eta^{2})^\ast$ and $\widetilde{h}(t,y,z)=(\widetilde{h}^1(t,y,z),\widetilde{h}^2(t,y,z))^\ast$ where
\[
\widetilde{h}^1(t,y,z):=h^{1}(t,y_1+y_2,y_2,z_1+z_2,z_2)-h^2(t,y_1+y_2,y_2,z_1+z_2,z_2)
\]
and
\[
\widetilde{h}^2(t,y,z):=h^2(t,y_1+y_2,y_2,z_1+z_2,z_2).
\]
Then statement (i) is equivalent to the following condition: \hfill \break
(iii) for any date $ U\in [0,T]$ and an arbitrary $\widetilde{\eta} = (\widetilde{\eta}^1,\widetilde{\eta}^2)$ such that $\widetilde{\eta}^1\ge 0$, the unique solution $(\widetilde{Y},\widetilde{Z})$ to the following BSDE over time interval $[0,U]$
\be \label{transferred two dimensioanl Brownian BSDE}
\widetilde{Y}_{t}=\widetilde{\eta}+\int_{t}^{U}\widetilde{h}(s,\widetilde{Y}_s, \widetilde{Z}_s)\, ds-\int_{t}^{U }\widetilde{Z}_s\, dW_s
\ee
satisfies $\widetilde{Y}^1\ge 0$. By applying Proposition \ref{BSVP proposition} to BSDE (\ref{transferred two dimensioanl Brownian BSDE}) and the convex, closed set $K = \mathbb{R}_{+}\times\mathbb{R}$, we see that (iii) is in turn equivalent to (ii),
since \eqref{bsvp1} coincides with \eqref{BSVP condition} in that case.
\end{proof}

\subsection{Initial Endowments of Equal  Signs}  \label{sec4.3}

In Sections \ref{sec4.3} and \ref{sec4.4}, we work under Assumption \ref{assdif}, so that we deal with a diffusion model.
For simplicity, we present here the case of one risky asset driven by the one-dimensional Brownian motion $W$ but, in view of Theorem
\ref{comparison theorem for coupled BSDE}, an extension to the case of $d$ risky assets driven by a $d$-dimensional Brownian motion is rather straightforward. Let us recall that the process $a$ is given by equation \eqref{defaa}.

\bhyp \label{assumption for stock coefficients}
We postulate that the process $a$ satisfies Novikov's condition (\ref{Novikov condition}), the process $(\sigma(\cdot, S))^{-1}$ and all the interest rates are continuous processes, and the process $(\sigma(\cdot, S))^{-1}S$ is bounded.
\ehyp

Since
\bde
d\wt S^{l,\textrm{cld}}_t =
\big( \mu(t, S_{t})+\kappa(t, S_{t})-\rll_t S_{t}\big) \, dt + \sigma (t, S_{t})\, dW_t = \sigma (t, S_{t}) (a_t \, dt + dW_t),
\ede
the pricing BSDE (\ref{main two dimensioanl BSDE}) reduces to
\bde
\left\{ \begin{array}
[c]{l}
dY_t = Z_t\sigma (t, S_{t})\, dW_t + \big( g(t,Y_t, Z_t)+\sigma (t, S_{t}) a_t Z_t \big)\, dt + d\overline{A}_t, \medskip\\
Y_T=0,
\end{array} \right.
\ede
or, equivalently,
\be \label{main two dimensioanl Brownian BSDE with general A}
\left\{ \begin{array}
[c]{l}
dY_t = Z_t\, dW_t
+ \big( g \big(t,Y_t,(\sigma (t, S_{t}))^{-1}Z_t \big)+a_t Z_t \big)\, dt + d\overline{A}_t, \medskip\\
Y_T=0.
\end{array} \right.
\ee
We first focus on the valuation and hedging of the collateralized European contingent claim $(H_T,C)$
given by \eqref{euro}. Then (\ref{main two dimensioanl Brownian BSDE with general A}) is equivalent
to the following BSDE, for $t \in [0,T)$,
\bde
Y_t =\begin{pmatrix}
-H_T \\ -H_T
\end{pmatrix}-\int_{t}^{T}Z_s\, dW_s
-\int_{t}^{T} \big(g (s,Y_s, (\sigma(s, S_{s}))^{-1} Z_s )+a_s Z_s \big)\, ds
\ede
with an additional jump at terminal date $T$, which ensures that $Y_T=0$.  It is thus clear that it suffices to examine the following BSDE on $[0,T]$
\be \label{main two dimensioanl Brownian BSDEx}
\left\{ \begin{array}
[c]{l}
dY_t = Z_t\, d\wtWl_t + g \big(t,Y_t, (\sigma(t, S_{t}))^{-1}Z_t \big) \, dt, \medskip\\
Y_T=(-H_T,-H_T)^{\ast},
\end{array} \right.
\ee
where $\wtWl $ is a Brownian motion under the probability measure $\PT^l$ defined by \eqref{defptl}.
We are now in a position to study the range of fair bilateral prices at time $t$ for the European claim with negotiated collateral.
Recall that in the present framework we have that $P^{h}(x_{1},A,C) =P^{h}(x_{1},x_2,A,C)$ and $P^{c}(x_2,-A,-C)= P^{c}(x_{1},x_2,-A,-C)$.

\bp \label{general Bergman model inequality proposition for both positive initial wealth}
Let $x_{1}\ge0,\, x_{2}\ge0$ and Assumptions \ref{assdif}, \ref{hypo2} and \ref{assumption for stock coefficients} be valid. Consider an arbitrary collateralized European claim $(H_T,C)$ where $H_T \in L^{2}(\Omega,\mathcal{F}_T,\PT^l )$.  Then we have, for every $t\in[0,T]$,
\be \label{range3}
P^{c}_t (x_{2},-H_T ,-C)\leq P^{h}_t (x_{1},H_T,C),  \quad \PT^l-\aass
\ee
\ep

\begin{proof}
We write $\sigma^{-1}:=(\sigma(t, S_{t}))^{-1}$. It is sufficient to check that the functions $h^1$ and $h^2$, which are given by
\[
h^{1}(t,y_1,y_2,z_1,z_2):=-g^{1}\big(t,y_1,y_2,\sigma^{-1}z_1,\sigma^{-1}z_2 \big)
\]
and
\[
h^{2}(t,y^1,y_2,z_1,z_2):=-g^{2}\big(t,y_1,y_2,\sigma^{-1}z_1,\sigma^{-1}z_2 \big),
\]
where $g^{1}$ and $g^{2}$ are given by (\ref{generator 1}) and (\ref{generator 2}) with $d=1$, respectively,
satisfy Assumption  \ref{assumption for Brownian BSDE generator} under $\PT^l$ and condition (\ref{BSVP condition}).
First, using the continuity of $(\sigma(\cdot, S))^{-1}$, $g^{1}$ and $g^{2}$ with respect to $t$, we know that for $y_1,y_2,z_1,z_2\in\mathbb{R}$, the function $h^{1}(t,y_1,y_2,z_1,z_2)$ and $h^{2}(t,y_1,y_2,z_1,z_2)$ are also continuous with respect to $t$. Second, since the process $(\sigma(\cdot, S))^{-1}S$ is bounded and the function $\Cnq $ is uniformly Lipschitz continuous, it is obvious that $h^{1}(t,y_1,y_2,z_1,z_2)$ and $h^{2}(t,y_1,y_2,z_1,z_2)$ are uniformly Lipschitz continuous with respect to $(y_1,y_2,z_1,z_2)$. Moreover, from $\Cnq (0,0)=0$ and $x_1,x_2\ge0$, we obtain
\[
h^{1}(t,0,0,0,0)=h^{2}(t,0,0,0,0)=0.
\]
We conclude that Assumption  \ref{assumption for Brownian BSDE generator} holds for $h^{1}$ and $h^{2}$.
Let us check condition (\ref{BSVP condition}) is valid as well. If we set
\[
\delta_1:=y_1^++y_2+\Cnq (-y_1^+-y_2,-y_2)+x_1B_{t}^{l}-(B_{t}^{l})^{-1}\sigma^{-1}(z_1+z_2) S_t
\]
and
\[
\delta_2:=-y_2-\Cnq (-y_1^+-y_2,- y_2)+x_2B_{t}^{l}+(B_{t}^{l})^{-1}\sigma^{-1}z_2 S_t,
\]
then
\[
\begin{array}[c]{rl}
&h^{1}(t,y_1^++y_2,y_2,z_1+z_2,z_2)-h^{2}(t,y_1^++y_2,y_2,z_1+z_2,z_2)\medskip\\
&=-g^{1}(t,y_1^++y_2,y_2,\sigma^{-1}(z_1+z_2),\sigma^{-1}z_2)+g^{2}(t,y_1^++y_2,y_2,\sigma^{-1}(z_1+z_2),\sigma^{-1}z_2)\medskip\\
&=-r_{t}^{l}(B_t^{l})^{-1}\sigma^{-1}z_1 S_t +(x_1+x_2) B_{t}^{l}r_{t}^{l}-r_{t}^{l}(\delta_1^++\delta_2^+)+r_{t}^{b}(\delta_1^-+\delta_2^-).
\end{array}
\]
Since $r_{t}^{l}\leq r_{t}^{b}$, we have
\[
\begin{array}[c]{rl}
r_{t}^{l}(\delta_1^++\delta_2^+)-r_{t}^{b}(\delta_1^-+\delta_2^-)& \leq r_{t}^{l}(\delta_1^++\delta_2^+)-r_{t}^{l}(\delta_1^-+\delta_2^-) = r_{t}^{l}(\delta_1+\delta_2)\medskip\\
&= r_{t}^{l}y_1^++ (x_1+x_2) B_{t}^{l}r_{t}^{l}-r_{t}^{l}(B_t^{l})^{-1}\sigma^{-1}z_1 S_t
\end{array}
\]
and
\[
\begin{array}[c]{rl}
&h^{1}(t,y_1^++y_2,y_2,z_1+z_2,z_2)-h^{2}(t,y_1^++y_2,y_2,z_1+z_2,z_2)\medskip\\
&=-r_{t}^{l}(B_t^{l})^{-1}\sigma^{-1}z_1 S_t +(x_1+x_2) B_{t}^{l}r_{t}^{l}-r_{t}^{l}(\delta_1^++\delta_2^+)+r_{t}^{b}(\delta_1^-+\delta_2^-)
\ge-r_{t}^{l}y_1^+.
\end{array}
\]
Consequently,  we obtain
\[
\begin{array}[c]{rl}
& -4y_1^-[ h^{1}(t,y_1^++y_2,y_2,z_1+z_2,z_2)-h^{2}(t,y_1^++y_2,y_2,z_1+z_2,z_2)]\medskip\\
&\leq  4r_{t}^{l}y_1^-y_1^+=0\leq  M|y_1^-|^2+2   z_1^2 \I_{\{y_1<0\}},
\end{array}
\]
which is the desired condition (\ref{BSVP condition}).
\end{proof}

Let us now consider a more general contract $A$ where the hedger receives cash flows $H_1,H_2,\ldots,H_k$ at times $0< t_1\leq t_2\leq\ldots\leq t_k\leq T$, so that
\[
A_t-A_0= \sum_{l=1}^{k} \I_{[t_l,T]}(t)H_l
\]
where $H_l\in L^{2}(\Omega,\mathcal{F}_{t_l}, \PT^l )$. For conciseness, we denote this claim as $(H,C)$.

\bp \label{proposition for special contract 1}
Let $x_{1}\ge0,\, x_{2}\ge0$ and Assumptions \ref{assdif}, \ref{hypo2} and \ref{assumption for stock coefficients} be valid. Then for any collateralized claim $(H,C)$ where $H_l\in L^{2}(\Omega,\mathcal{F}_{t_l}, \PT^l )$ for $l=1,2,\dots ,k$ we have, for every $t\in[0,T]$,
\bde 
P^{c}_t (x_{2},-H,-C)\leq P^{h}_t (x_{1},H,C),  \quad \PT^l -\aass
\ede
\ep

\begin{proof}
We first study the problem on $[t_{k},T]$. Since $dA_t =0$, it is just a special case of Proposition \ref{general Bergman model inequality proposition for both positive initial wealth} (it suffices to take $H_T =0$), we have that $P^{c}_t \leq P^{h}_t$ for all $t\in [t_k,T]$.
Indeed, one can check directly that the equalities $P^{c}_t =P^{h}_t=0$ hold for all $t\in [t_k,T]$.

We now consider the problem on $[t_{k-1},t_k)$. Recall that $(P^h,P^c)^\ast=Y=(Y_1, Y_2)^\ast$ where $(Y,Z)$ solves BSDE (\ref{main two dimensioanl Brownian BSDE with general A}). From the first step, we know that $Y_{1,t_k}=Y_{2,t_k}=0$. Let us consider  BSDE (\ref{main two dimensioanl Brownian BSDE with general A}) on $[t_{k-1},t_k]$. Noticing that  $A$ only changes at time $t_k$ and $\Delta A_{t_k}=H_k$, we obtain, for $s\in [t_{k-1},t_k)$,
\bde
Y_s =\begin{pmatrix}
-H_k \\
-H_k
\end{pmatrix}-\int_{s}^{t_k}Z_t\, d\wtWl_t -\int_{s}^{t_k}g \big(t,Y_t, \sigma^{-1}_t Z_t \big)\, dt
\ede
where $\sigma^{-1}_t :=(\sigma(t, S_{t}))^{-1}$.
So this is nothing else than just the pricing BSDE for European claim with maturity $t_k$ and with receiving payoff $H_k$. Therefore, using Proposition \ref{general Bergman model inequality proposition for both positive initial wealth}, we have that  for all $t\in[t_{k-1},t_k)$, $Y_{2,t}\leq Y_{1,t}$ which yields $P^{c}_t \leq P^{h}_t$.

We can extend this inequality to $[t_{k-2},t_{k-1})$. Indeed,  for $s\in [t_{k-2},t_{k-1})$,
\bde
Y_s =\begin{pmatrix}
Y_{1,t_{k-1}}-H_{k-1} \\
Y_{2,t_{k-1}}-H_{k-1}
\end{pmatrix}-\int_{s}^{t_{k-1}}Z_t\, d\wtWl_t
-\int_{s}^{t_{k-1}} g \big(t,Y_t, \sigma^{-1}_tZ_t \big)\, dt.
\ede
Since from the second step, we know $Y_{2,t_{k-1}}\leq Y_{1,t_{k-1}}$, using Theorem \ref{comparison theorem for coupled BSDE} and the proof of Proposition \ref{general Bergman model inequality proposition for both positive initial wealth}, we obtain
$Y_{2,t}\leq Y_{1,t}$ for all $t\in[t_{k-2},t_{k-1})$, which in turn yields $P^{c}_t \leq P^{h}_t$ for all $t\in[t_{k-2},t_{k-1})$.
By the backward induction, we conclude that \eqref{range3} holds for every $t\in[0,T]$.
\end{proof}

 We also have the following result for the counterparty's price.

\bp \label{proposition for special contract 2}
Let $x_{1}\ge0,\, x_{2}\ge0$ and Assumptions \ref{assdif}, \ref{hypo2} and \ref{assumption for stock coefficients} be valid. Consider
an arbitrary contract $(A,C)$ where $A - A_0$ is a non-positive (or bounded from above, so that $A - A_0\leq M$ for some constant $M$), continuous, $\gg$-adapted process such that $\mathbb{E}_{\widetilde{\mathbb{P}}^l}[\sup_{t\in[0,T]}|A_t|^2]<\infty$. Then we have,
 for every $t\in[0,T]$,
\bde  
P^{c}_t (x_{2},-A,-C)\leq P^{h}_t (x_{1},A,C),  \quad \PT^l-\aass
\ede
\ep

\begin{proof}
 Recall that $\sigma^{-1}_t:=(\sigma(t, S_{t}))^{-1}$. We have $(P^h,P^c)^\ast=Y=(Y_1, Y_2)^\ast$ where $(Y,Z)$ solves BSDE (\ref{main two dimensioanl Brownian BSDE with general A}). Let $\widetilde{Y}:=Y-\overline{A}+\overline{A}_0$, where $\overline{A}=(A,A)^{\ast}$ and $\overline{A}_0=(A_0,A_0)^{\ast}$, so that
\bde
\left\{ \begin{array}
[c]{l}
d\widetilde{Y}_t = Z_t\, d\wtWl_t
+g \big(t,\widetilde{Y}_t+\overline{A}_t-\overline{A}_0, \sigma^{-1}_tZ_t \big) \, dt, \medskip\\
\widetilde{Y}_T=-\overline{A}_T.
\end{array} \right.
\ede
Similarly as in the proof of Proposition \ref{general Bergman model inequality proposition for both positive initial wealth}, we let
\[
h^{1}(t,y_1,y_2,z_1,z_2):=-g^{1}(t,y_1+A_t-A_0,y_2+A_t-A_0,\sigma^{-1}_tz_1,\sigma^{-1}_tz_2)
\]
and
\[
h^{2}(t,y^1,y_2,z_1,z_2):=-g^{2}(t,y_1+A_t-A_0,y_2+A_t-A_0,\sigma^{-1}_tz_1,\sigma^{-1}_tz_2).
\]
Since $A$ is continuous and $\mathbb{E}_{\widetilde{\mathbb{P}}^l}[\sup_{t\in[0,T]}|A_t|^2]<\infty$, it is not hard to check that Assumption  \ref{assumption for Brownian BSDE generator} is satisfied by $h^1$ and $h^{2}$. Moreover, since $A-A_0\leq 0$ (or $A-A_0\leq M$), we have
\[
\begin{array}[c]{rl}
& -4y_1^-[ h^{1}(t,y_1^++y_2,y_2,z_1+z_2,z_2)-h^{2}(t,y_1^++y_2,y_2,z_1+z_2,z_2)]\medskip\\
& \leq 4r_{t}^{l}y_1^-(y_1^++A_t-A_0)\leq |M||y_1^-|^2+2   z_1^2 \I_{\{y_1<0\}}.
\end{array}
\]
To complete the proof, it suffices to use Theorem \ref{comparison theorem for coupled BSDE}.
\end{proof}

\brem
For a contract $(A,C)$ with a more general process $A$, we may not have similar results. This is because that in (\ref{main two dimensioanl Brownian BSDE with general A}), a general cash flow $\overline{A}$ may destroy the viability property.
However, by mixing the two kinds of special contracts introduced in Propositions \ref{proposition for special contract 1} and \ref{proposition for special contract 2}, we can construct the following class of contracts: for $0<t_1\leq t_2\leq\ldots\leq t_k\leq T$, and processes $H_l(t),\, l=1,\dots , k$ defined on $[t_l,T]$,
\[
A_t-A_0=\sum_{l=1}^{k} \I_{[t_l,T]}(t)H_l(t)
\]
where, for $l=1,2,\ldots,k$, the process $H_{l}(t),\, t \in [t_l,T]$, satisfies one of the following conditions: \hfill \break
(i) $H_{l}$ is a continuous, $\gg$-adapted process, $H_{l}\leq M$ and $\mathbb{E}_{\widetilde{\mathbb{P}}^l}[\sup_{t\in[t_l,T]}|H_{l}(t)|^2]<\infty$, \hfill \break
(ii) $H_{l}(t)=H_l$ for all $t\in[t_l,T]$, where the random variable $H_l\in L^{2}(\Omega,\mathcal{F}_{t_l},\mathbb{P})$.

By combining the statements and proofs of Propositions \ref{proposition for special contract 1} and \ref{proposition for special contract 2}, one can show that the range of fair bilateral prices ${\cal R}^f_t(x_1,x_2)$ for the contract
 $(A,C)$ satisfying (i)--(ii) is non-empty almost surely.
\erem

\subsection{Initial Endowments of Opposite Signs}  \label{sec4.4}

We only consider here the case of a collateralized European contingent claim $(H_T,C)$, but similar results hold for two special kinds of contracts introduced in Propositions \ref{proposition for special contract 1} and \ref{proposition for special contract 2}.
We work under Assumptions \ref{assdif} and \ref{changed assumption for artifical cumulative dividend price} and we denote
\be \label{defbb}
b_{t}:=(\sigma(t, S_{t}))^{-1}\big( \mu(t, S_{t})+\kappa(t, S_{t})-\beta_t S_{t} \big).
\ee

\bhyp \label{assumption 2 for stock coefficients}
We postulate  that the process $b$ satisfies Novikov's condition (\ref{Novikov condition}), the processes $(\sigma(\cdot, S))^{-1}, \beta$ and all interest rates are continuous processes and the process $(\sigma(\cdot, S))^{-1}S$ is bounded.
\ehyp

We observe that
\bde
d\wt S^{\textrm{cld}}_t =
\big( \mu(t, S_{t})+\kappa(t, S_{t})-\beta_t S_{t}\big) \, dt + \sigma (t, S_{t})\, dW_t =\sigma (t, S_{t})
\big( b_t\, dt+ dW_t \big)=\sigma (t, S_{t})\, d\wtWb_t
\ede
where $d\wtWb_{t}:=dW_{t}+b_{t}\, dt $. Let us define the probability measure $\PTb$ by setting
\bde
\frac{d\PTb}{d\P}=\exp\left\{-\int_{0}^{T}b_{t}\, dW_{t}
-\frac{1}{2}\int_{0}^{T}|b_{t}|^{2}\, dt \right\}.
\ede
From the Girsanov theorem, the process $\wtWb$ is the Brownian motion under $\PTb$ and thus $\wt S^{\textrm{cld}}$ is a $(\PTb , \gg)$-(local) martingale with the quadratic variation $\langle \wt S^{\textrm{cld}}\rangle_t=\int_{0}^t |\sigma(u, S_{u})|^{2}\, du$. Moreover, since the process $(\sigma(\cdot, S))^{-1}S$ is bounded,  Assumption \ref{changed assumption for artifical cumulative dividend price} holds.
We conclude that the model is arbitrage free under $\PTb$ (see Proposition \ref{remark for non-arbitrage model}).

Under the present framework, BSDE (\ref{main two dimensioanl BSDE 2}) can be represented as follows
\bde
\left\{ \begin{array}
[c]{l}
dY_t = Z_t\sigma (t, S_{t})\, dW_t
+\big(\widehat{g }\big(t,Y_t, Z_t \big)+\sigma (t, S_{t}) b_t Z_t \big)\, dt + d\overline{A}_t, \medskip\\
Y_T=0 .
\end{array} \right.
\ede
As in  Section \ref{sec4.3}, it is thus sufficient to examine the following BSDE on $[0,T]$
\bde
\left\{ \begin{array}
[c]{l}
dY_t = Z_t\, d\wtWb_t +\widehat{g} \big(t,Y_t, (\sigma(t, S_{t}))^{-1}Z_t \big)\, dt, \medskip\\
Y_T=(-H_T,-H_T)^{\ast}.
\end{array} \right.
\ede

\bp \label{general Bergman model inequality proposition for opposite initial wealth}
Let $x_{1}\ge0,\, x_{2}\leq 0$ be such that $x_1x_2=0$. If Assumptions  \ref{assdif}, \ref{changed assumption for artifical cumulative dividend price}, \ref{hypo2} and \ref{assumption 2 for stock coefficients} are satisfied, then
for any collateralized European claim $(H_T,C)$ such that $H_T \in L^{2}(\Omega,\mathcal{F}_T, \PTb )$ we have, for every $t\in[0,T]$,
\bde
P^{c}_t (x_{2},-H_T,-C)\leq P^{h}_t (x_{1},H_T,C),  \quad\PTb-\aass
\ede
\ep

\begin{proof}
Let $\sigma^{-1}_t:=(\sigma(t, S_{t}))^{-1}$. It suffices to check that the functions
\[
h^{1}(t,y_1,y_2,z_1,z_2):=-\widehat{g}^{1}\big(t,y_1,y_2,\sigma^{-1}_tz_1,\sigma^{-1}_tz_2 \big)
\]
and
\[
h^{2}(t,y^1,y_2,z_1,z_2):=-\widehat{g}^{2}\big(t,y_1,y_2,\sigma^{-1}_tz_1,\sigma^{-1}_tz_2\big)
\]
satisfy Assumption  \ref{assumption for Brownian BSDE generator} and condition (\ref{BSVP condition}), where $\widehat{g}^{1}$ and $\widehat{g}^{2}$ are given by (\ref{generator 3}) and (\ref{generator 4}) with $d=1$, respectively.
First, from the continuity of $\beta$, $\sigma^{-1}$, $\widehat{g}^{1}$ and $\widehat{g}^{2}$ with respect to $t$, we deduce that for $y_1,y_2,z_1,z_2\in\mathbb{R}$, the functions $h^{1}(t,y_1,y_2,z_1,z_2)$ and $h^{2}(t,y_1,y_2,z_1,z_2)$ are also continuous with respect to $t$. Second,  since  $\sigma^{-1}S$ is bounded and $\Cnq $ is uniformly Lipschitz continuous, it is clear that $h^{1}(t,y_1,y_2,z_1,z_2)$ and $h^{2}(t,y_1,y_2,z_1,z_2)$ are uniformly Lipschitz continuous with respect to $(y_1,y_2,z_1,z_2)$. Moreover, from $\Cnq (0,0)=0$ and $x_1\ge0$, and $x_2\leq0$, we have that $h^{1}(t,0,0,0,0)=h^{2}(t,0,0,0,0)=0$.
We thus see that Assumption  \ref{assumption for Brownian BSDE generator} holds for $h^{1}$ and $h^{2}$. Finally, let us check that condition (\ref{BSVP condition}) is met as well. To this end, we set
\[
\delta_1:=y_1^++y_2+\Cnq (-y_1^+-y_2,-y_2)+x_1B_{t}^{l}-\sigma^{-1}_t(z_1+z_2) S_t
\]
and
\[
\delta_2:=-y_2-\Cnq (-y_1^+-y_2,-y_2)+x_2B_{t}^{b}+\sigma^{-1}_tz_2 S_t.
\]
Then
\[
\begin{array}[c]{rl}
&h^{1}(t,y_1^++y_2,y_2,z_1+z_2,z_2)-h^{2}(t,y_1^++y_2,y_2,z_1+z_2,z_2)\medskip\\
&=-\widehat{g}^{1}(t,y_1^++y_2,y_2,\sigma^{-1}_t(z_1+z_2),\sigma^{-1}_tz_2)+\widehat{g}^{2}(t,y_1^++y_2,y_2,\sigma^{-1}_t(z_1+z_2),\sigma^{-1}_tz_2)\medskip\\
&=-\sigma^{-1}_t\beta_{t}(z_1+z_2) S_t +x_1 B_{t}^{l}r_{t}^{l}+r_{t}^{c}\Cnq (-y_1^+-y_2,-y_2)-r_{t}^{l}\delta_1^++r_{t}^{b}\delta_1^-\medskip\\
&\mbox{}\ \ +\sigma^{-1}_t\beta_{t}z_2S_t +x_2 B_{t}^{b}r_{t}^{b}-r_{t}^{c}\Cnq (-y_1^+-y_2,-y_2)-r_{t}^{l}\delta_2^++r_{t}^{b}\delta_2^-\medskip\\
&=-\sigma^{-1}_t\beta_{t}z_1S_t+x_1B_{t}^{l}r_{t}^{l}+x_2 B_{t}^{b}r_{t}^{b}-r_{t}^{l}(\delta_1^++\delta_2^+)+r_{t}^{b}(\delta_1^-+\delta_2^-).
\end{array}
\]
Since $r_{t}^{l}\leq r_{t}^{b}$, we have
\[
\begin{array}[c]{rl}
 & r_{t}^{l}(\delta_1^++\delta_2^+)-r_{t}^{b}(\delta_1^-+\delta_2^-)\leq  \min\left\{ r_{t}^{l}(\delta_1+\delta_2),  r_{t}^{b}(\delta_1+\delta_2)\right\}\medskip\\
& = \min\left\{ r_{t}^{l}y_1^++x_1B_{t}^{l}r_{t}^{l}+x_2 B_{t}^{b}r_{t}^{l}-r_{t}^{l}\sigma^{-1}_tz_1 S_t,\, r_{t}^{b}y_1^++x_1B_{t}^{l}r_{t}^{b}+x_2 B_{t}^{b}r_{t}^{b}-r_{t}^{b}\sigma^{-1}_tz_1 S_t\right\}.
\end{array}
\]
Thus
\[
\begin{array}[c]{rl}
&h^{1}(t,y_1^++y_2,y_2,z_1+z_2,z_2)-h^{2}(t,y_1^++y_2,y_2,z_1+z_2,z_2) \ge-\sigma^{-1}_t\beta_{t}z_1S_t\medskip\\
&+\max\left\{ -r_{t}^{l}y_1^++x_2B_{t}^{b}r_{t}^{b}-x_2 B_{t}^{b}r_{t}^{l}+r_{t}^{l}\sigma^{-1}_tz_1 S_t,\, -r_{t}^{b}y_1^++x_1B_{t}^{l}r_{t}^{l}-x_1B_{t}^{l}r_{t}^{b}+r_{t}^{b}\sigma^{-1}_tz_1 S_t\right\}.
\end{array}
\]
We also have that
\[
\begin{array}[c]{rl}
&h^{1}(t,y_1^++y_2,y_2,z_1+z_2,z_2)-h^{2}(t,y_1^++y_2,y_2,z_1+z_2,z_2)\medskip\\
&\ge -r_{t}^{l}y_1^++\sigma^{-1}_tS_t(r_{t}^{l}-\beta_{t})z_1+x_2B_{t}^{b}(r_{t}^{b}-r_{t}^{l}).
\end{array}
\]
Consequently, if $x_2=0$ then, using the boundedness of processes $\beta$, $r^l$ and $\sigma^{-1}S$, we obtain
\[
\begin{array}[c]{rl}
& -4y_1^-[ h^{1}(t,y_1^++y_2,y_2,z_1+z_2,z_2)-h^{2}(t,y_1^++y_2,y_2,z_1+z_2,z_2)]\medskip\\
&\leq  4r_{t}^{l}y_1^-y_1^+-4y_1^- z_1\sigma^{-1}_tS_t(r_{t}^{l}-\beta_{t}) =-4y_1^- z_1\sigma^{-1}_tS_t(r_{t}^{l}-\beta_{t})\medskip\\
&\leq  M|y_1^-|^2+2   z_1^2 \I_{\{y_1<0\}},
\end{array}
\]
which is the desired inequality (\ref{BSVP condition}).  The same inequality can be obtained when $x_1=0$.
\end{proof}

\brem
Let us consider a more general class of contracts considered in Section \ref{sec4.3}.
We now assume that the interest rates $r^{l}$ and $r^{b}$ are deterministic and satisfy $r^{l}_{t}<r^{b}_{t}$ for all $t\in[0,T]$. Using the example studied in the proof of Proposition 5.4 in \cite{NR2}, for every model for risky assets we see that for every contract $(A,C)$ considered in Section \ref{sec4.3}, for all $t\in[0,T]$,
\bde
P^{c}_t (x_{2},-A,-C)\leq P^{h}_t (x_{1},A,C),  \quad \PTb-\aass
\ede
if and only if $x_{1}x_{2}=0$.
\erem

\brem
As was explained in Section 3.1, the price $P^h(x_1,A,C)$ (resp., $P^c(x_2,-A-,C)$) indeed should be $P^h(x_1,x_2,A,C)$ (resp., $P^c(x_1,x_2,-A,-C)$), meaning that the hedger's and the counterparty's price depend on  both initial endowments $x_1$ and $x_2$. Using the comparison theorem for multi-dimensional BSDE (see Hu and Peng \cite{HP-2006}), one may attempt to show the monotonicity of prices with respect to the initial endowment (for related results, see Section 5.4 in \cite{NR2}).
\erem

%

\section{Model with Partial Netting and Hedger's Collateral} \label{sec5}

In Sections \ref{sec5} and \ref{sec6}, we consider the model with partial netting and full rehypothecation of the cash collateral. For a detailed description of this modeling framework, the reader is referred to \cite{BR-2014,NR2}. Our aim is to show that the methodology
developed in preceding sections can be applied to this setup, albeit with possibly different conclusions regarding the properties
of unilateral and bilateral prices. Since the proofs of some results are very similar to the proofs of their counterparts in
Bergman's model, they are omitted.

 From Lemma 2.1 and Lemma 2.2 of \cite{NR2},  we know that for a self-financing trading strategy
\[
\varphi=(\xi^1, \dots , \xi^d ,\varphi^l,\varphi^b,\varphi^{1,b},\varphi^{2,b},\ldots,\varphi^{d,b},\eta),
 \]
the processes $Y^{l}:=(B^{l})^{-1}V^p(x,\phi , \pA ,\pC)$ and $Z^{l,i}=\xi^i ,\, i=1,2,\ldots,d$ satisfy
\be  \label{partial netting model lending BSDE}
dY_{t}^{l} = \sum_{i=1}^dZ^{l,i}_t \, d\wt S^{i,l,{\textrm{cld}}}_t+G_{l}(t,Y_{t}^{l},Z_{t}^{l})\, dt+ dA^{C,l}_t
\ee
where the generator $G_l$ equals, for all $(\omega , t,y,z)\in \Omega \times [0,T] \times \mathbb{R}\times\mathbb{R}^{d}$,
\bde
\begin{array}
[c]{rl}
G_{l}(t,y,z)=&(B_{t}^{l})^{-1}\sum_{i=1}^d r_{t}^{l} z^i S^i_t-(B_{t}^{l})^{-1}\sum_{i=1}^d r_{t}^{i,b}(z^i S^i_t)^{+}-r_{t}^{l}y\medskip\\
&+(B_{t}^{l})^{-1}\bigg(r_{t}^{l}\Big(yB_{t}^{l}+\sum_{i=1}^d (z^i S^i_t)^-\Big)^+-r_{t}^{b}\Big(yB_{t}^{l}+\sum_{i=1}^d (z^i S^i_t)^-\Big)^- \bigg).
\end{array}
\ede
Similarly, the processes $Y^{b}:=(B^{b})^{-1}V^p (x,\phi , \pA ,\pC)$ and $Z^{b,i}=\xi^i,\, i=1,2,\ldots,d$ satisfy
\be \label{partial netting model borrowing BSDE}
dY_{t}^{b} = \sum_{i=1}^dZ^{b,i}_t \, d\wt S^{i,b,{\textrm{cld}}}_t+G_{b}(t,Y_{t}^{b},Z_{t}^{b})\, dt+ dA_t^{C,b}
\ee
where, for all $(\omega ,t,y,z)\in \Omega \times [0,T] \times \mathbb{R}\times\mathbb{R}^{d}$,
\bde
\begin{array}
[c]{rl}
G_{b}(t,y,z)=&(B_{t}^{b})^{-1}\sum_{i=1}^d r_{t}^{b}z^i S^i_t-(B_{t}^{b})^{-1}\sum_{i=1}^d r_{t}^{i,b}(z^i S^i_t)^{+}-r_{t}^{b}y\medskip\\
&+(B_{t}^{b})^{-1}\bigg(r_{t}^{l}\Big(yB_{t}^{b}+\sum_{i=1}^d (z^i S^i_t)^-\Big)^+-r_{t}^{b}\Big(yB_{t}^{b}+\sum_{i=1}^d (z^i S^i_t)^-\Big)^-\bigg).
\end{array}
\ede
Throughout Section \ref{sec5}, we work under Assumption \ref{hypo1} of hedger's collateral.



\subsection{Initial Endowments of Equal Signs}  \label{sec5.5}

We first examine the case where the initial endowments satisfy $x_{1}\ge0$ and $x_{2}\ge0$. We notice that in such case, under Assumption \ref{assumption for lending cumulative dividend price} (or Assumption \ref{changed assumption for lending cumulative dividend price}) the partial netting model is arbitrage-free with respect to any contract $(A,C)$ for the hedger and the counterparty (see Proposition 3.1 in \cite{NR2}). Using Propositions 4.1 and 4.2 in \cite{NR2}, we can establish the following proposition, which corresponds to Proposition \ref{hedger collateral dependent Bergman model ex-dividend price}  in the present work.

\bp \label{hedger collateral dependent partial netting model ex-dividend price}
Let $x_1 \geq 0,\, x_2\ge0$ and Assumptions \ref{hypo1} and \ref{changed assumption for lending cumulative dividend price} be valid.
For any contract $(A,C)$ such that $A \in \Classapl $, the hedger's ex-dividend price equals $P^h:=P^{h}(x_{1},A,C) = Y^1$ where $(Y^{1}, Z^{1})$ is the unique solution to the BSDE
\be \label{partial netting model main BSDE hedger}
\left\{ \begin{array}
[c]{l}
dY_t^1 = Z^{1,\ast}_t \, d \wt S^{l,{\textrm{cld}}}_t
+f_l \big(t, x_1,Y^1_t, Z^1_t \big)\, dt + dA_t, \medskip\\
Y^1_T=0,
\end{array} \right.
\ee
with the generator $f_{l}$ given by
\be \label{partial netting model one value dependent generator for hedger}
\begin{array}
[c]{rl}
f_{l}(t,x_1,y,z)=& r_{t}^{l}(B_t^{l})^{-1}z^\ast S_t -(B_t^{l})^{-1}\sum_{i=1}^d r_{t}^{i,b}(z^i S^i_t)^{+}-x_1 B_{t}^{l}r_{t}^{l}-r_{t}^{c}q(-y)\medskip\\
&\mbox{}+r_{t}^{l}\Big(y+q(-y)+x_1B_{t}^{l}+(B_{t}^{l})^{-1}\sum_{i=1}^d (z^i S^i_t)^-\Big)^+\medskip\\
&\mbox{}-r_{t}^{b}\Big(y+q(-y)+x_1B_{t}^{l}+(B_{t}^{l})^{-1}\sum_{i=1}^d (z^i S^i_t)^-\Big)^-
\end{array}
\ee
and the counterparty's ex-dividend price equals $P^{c}:=P^{c} (x_2,-A,-C) =Y^2$ where $(Y^{2}, Z^{2})$ is the unique solution to the BSDE
\be \label{partial netting model main BSDE counterparty}
\left\{ \begin{array}
[c]{l}
dY_t^2 = Z^{2,\ast}_t \, d \wt S^{l,{\textrm{cld}}}_t
+g_l \big(t, x_2,Y^2_t, Z^2_t \big)\, dt + dA_t, \medskip\\
Y^2_T=0 ,
\end{array} \right.
\ee
with the generator $g_{l}$ given by
\be \label{partial netting model one value dependent generator for counterparty}
\begin{array}
[c]{rl}
g_{l}(t,x_2,y,z)=&r_{t}^{l}(B_t^{l})^{-1}z^\ast S_t +(B_t^{l})^{-1}\sum_{i=1}^d r_{t}^{i,b}(-z^i S^i_t)^{+}+x_2 B_{t}^{l}r_{t}^{l}-r_{t}^{c}q(-Y_t^1)\medskip\\
&\mbox{}-r_{t}^{l}\Big(-y-q(-Y_t^1)+x_2B_{t}^{l}+(B_{t}^{l})^{-1}\sum_{i=1}^d (-z^i S^i_t)^-\Big)^+\medskip\\
&\mbox{}+r_{t}^{b}\Big(-y-q(-Y_t^1)+x_2B_{t}^{l}+(B_{t}^{l})^{-1}\sum_{i=1}^d (-z^i S^i_t)^-\Big)^-.
\end{array}
\ee
\ep

As in Bergman's model, if the convention of hedger's collateral is postulated, then we have $P^h = P^{h}(x_{1},A,C)$ and $P^{c}= P^{c}(x_{1},x_2,-A,-C)$, but we still denote the counterparty's price as $P^{c}(x_2,-A,-C)$. We are in a position to study the range of fair bilateral prices.

\bp \label{general partial netting model inequality proposition for both positive initial wealth}
Let $x_1 \geq 0,\, x_2\ge0$ and Assumptions \ref{hypo1} and \ref{changed assumption for lending cumulative dividend price} be valid.
Then for any contract $(A,C)$ such that $A \in \Classapl $ we have, for every $t\in[0,T]$,
\be \label{new}
P^{c}_t (x_{2},-A,-C)\leq P^{h}_t (x_{1},A,C),  \quad \PT^l-\aass
\ee
\ep

\begin{proof}
It is enough to show that $g_l(t, x_{2}, Y^{1},Z^{1})\ge f_l(t,x_{1}, Y^{1},Z^{1})$, $\PT^l\otimes \Leb-\aaee$. We denote
\begin{align*}
\delta &:=g_l(t, x_{2}, Y^{1},Z^{1})-f_l (t, x_{1}, Y^{1},Z^{1}) \\
&=\rll_tB_{t}^{l} (x_{1}+x_{2})+(B_t^{l})^{-1}\sum_{i=1}^d r_{t}^{i,b}|Z^{1,i}_t S^i_t|-\rll_t (\delta_{1}^{+}+\delta_{2}^{+})+\rbb_t (\delta_{1}^{-}+\delta_{2}^{-})
\end{align*}
where
\bde
\delta_{1}:=-Y^{1}_t-q(-Y^{1}_t)+\Blr_t x_{2}+(\Blr_t)^{-1}\sum_{i=1}^d (-Z^{1,i}_t S^i_t)^-
\ede
and
\bde
\delta_{2}:=Y^{1}_t+q(-Y^{1}_t)+\Blr_t x_{1}+(\Blr_t)^{-1}\sum_{i=1}^d (Z^{1,i}_t S^i_t)^-.
\ede
Since $r^{l}\leq r^b$ and $\rll\leq r^{i,b}$, we obtain
\bde
\begin{array}
[c]{rl}
\delta\ge &\rll_tB_{t}^{l} (x_{1}+x_{2})+(B_t^{l})^{-1}\sum_{i=1}^d r_{t}^{i,b}|Z^{1,i}_t S^i_t|-\rll_t (\delta_{1}+\delta_{2})\medskip\\
\ge &(B_t^{l})^{-1}\sum_{i=1}^d (r_{t}^{i,b}-r^l)|Z^{1,i}_t S^i_t|\ge 0 ,
\end{array}
\ede
which completes the proof.
\end{proof}

\subsubsection{Model with an Uncertain Money Market Rate}  \label{sec5.4}

 We study here the case of initial endowments satisfying $x_1 \ge 0 $ and $x_2\ge0$, but results for the case where $x_1\ge0$ and $x_2\leq 0$ are similar. Let us select an arbitrary $\gg$-adapted interest rate process satisfying
\be \label{rere1}
\rir_t\in[r^{l}_{t},r^{b}_t] \ \ \text{ for every } t\in[0,T].
\ee
We now consider the market model with the single money market rate $r$ in which the hedger and counterparty have the same ex-dividend price  $\Prir$ independent of their respective initial endowments. The price $\Prir = Y$ can be found by solving the BSDE
\be \label{partial netting main BSDE linear}
\left\{ \begin{array}
[c]{l}
dY_t = Z^{\ast}_t \, d \wt S^{l,{\textrm{cld}}}_t
+f(t, Y_t, Z_t )\, dt + dA_t, \medskip\\
Y_T=0 ,
\end{array} \right.
\ee
where the generator $f$ equals
\bde
f(t,y,z)=r_{t}^l(B_{t}^{l})^{-1}z^\ast S_t-(B_{t}^{l})^{-1}\sumik_{i=1}^dr^{i,b}_t(z^{i}S^i_t)^{+}-(B_{t}^{l})^{-1}\sumik_{i=1}^d\rir_t(z^{i}S^i_t)^{-} -r_{t}^{c}q(-y)+\rir_t\big(y+q(-y)\big).
\ede

Similarly to Proposition \ref{link to economies with no differential rates}, we have the following result under Assumption \ref{hypo1}.

\bp \label{partial netting link to economies with no differential rates}
For any contract $(A,C)$ such that $A \in \Classapl $, the unique no-arbitrage price in the market model with the money market rate $\rir$ satisfies $\Prir \leq P^{h}(0,A,C)$,  $\PT^l-\aass$ If $x_1=x_2=0$ and in addition, the function $q$ in (\ref{collateral 1}) satisfies $(\rir_t-r_{t}^{c})(q(y_1)-q(y_2))\leq 0$ for all $y_1\ge y_2$, then also $P^{c}(0,-A,-C) \leq \Prir$, $\PT^l-\aass$
\ep

\subsection{Initial Endowments of Opposite Signs}   \label{sec5.2}

Let us now consider the case where $x_{1}\ge0$ and $x_{2}\leq 0$. We now postulate that $r^b\leq r^{i,b}$ and Assumption \ref{changed assumption for artifical cumulative dividend price} holds with $r^b\leq \beta^i\leq r^{i,b}$.
From Proposition 3.2 in \cite{NR2}, we know that the partial netting model is arbitrage-free
for both the hedger and the counterparty in respect of any contract $(A,C)$ and arbitrary initial endowments.
 Using Proposition 5.3 in \cite{NR2} and argument similar as in the proof of Proposition \ref{opposite sign hedger collateral dependent Bergman model ex-dividend price}, one can prove the following propositions.

\bp \label{opposite sign hedger collateral dependent partial netting model ex-dividend price}
Let $x_{1}\ge0,\,  x_{2}\leq0$ and Assumptions \ref{hypo1} and \ref{changed assumption for artifical cumulative dividend price} be valid.
 For any contract $(A,C)$ such that $A \in \Classaptb $, the hedger's ex-dividend price equals $P^h= \overline{Y}^1$ where the pair $(\overline{Y}^{1}, \overline{Z}^{1})$ is the unique solution to the BSDE
\be \label{partial netting main BSDE hedger 1}
\left\{ \begin{array}
[c]{l}
d\overline{Y}_t^1 = \overline{Z}^{1,\ast}_t \, d \wt S^{{\textrm{cld}}}_t
+\overline{f} \big(t, x_1,\overline{Y}^1_t, \overline{Z}^1_t \big)\, dt + dA_t, \medskip\\
\overline{Y}^1_T=0,
\end{array} \right.
\ee
where
\bde
\begin{array}[c]{rl}
\overline{f}(t,x_1,y,z)&=\sum_{i=1}^dz^{i}\beta^{i}_{t}S_{t}^{i}-\sumik_{i=1}^dr^{i,b}_t(z^{i}S^i_t)^{+}-x_1\rll_t\Blr_t-r_{t}^{c}q(-y)
\medskip\\
&\mbox{}+r_{t}^{l}\Big(y+q(-y)+x_1B_{t}^{l}+\sumik_{i=1}^d(z^{i}S^i_t)^{-}\Big)^+\medskip\\
&\mbox{}-r_{t}^{b}\Big(y+q(-y)+x_1B_{t}^{l}
+\sumik_{i=1}^d(z^{i}S^i_t)^{-}\Big)^-
\end{array}
\ede
and the counterparty's ex-dividend price equals $P^c= \overline{Y}^c$ where the pair $(\overline{Y}^{c}, \overline{Z}^{c})$ is the unique solution to the BSDE
\be \label{partial netting main BSDE counterparty 1}
\left\{ \begin{array}
[c]{l}
d\overline{Y}_t^2 = \overline{Z}^{2,\ast}_t \, d \wt S^{{\textrm{cld}}}_t
+\overline{g} \big(t, x_2,\overline{Y}^2_t, \overline{Z}^2_t \big)\, dt + dA_t, \medskip\\
\overline{Y}^2_T=0,
\end{array} \right.
\ee
where
\bde
\begin{array}[c]{rl}
\overline{g}(t,x_2,y,z)=&\sum_{i=1}^dz^{i}\beta^{i}_{t}S_{t}^{i}+\sumik_{i=1}^dr^{i,b}_t(-z^{i}S^i_t)^{+}
+x_2\rbb_t\Bbr_t-r_{t}^{c}q(-\overline{Y}^1_{t})\medskip\\
&\mbox{}-r_{t}^{l}\Big(-y-q(-\overline{Y}^1_{t})+x_2B_{t}^{b}+\sumik_{i=1}^d(-z^{i}S^i_t)^{-}\Big)^+\medskip\\
&\mbox{}+r_{t}^{b}\Big(-y-q(-\overline{Y}^1_{t})+x_2B_{t}^{b}+\sumik_{i=1}^d(-z^{i}S^i_t)^{-}\Big)^-.
\end{array}
\ede
\ep

The following result shows that the range of fair bilateral prices is non-empty provided that $x_1 x_2 =0$.
Otherwise, one can produce an example of a model in which this range is empty.

\bp \label{partial netting model inequality proposition for positive negative initial wealth}
Let $x_{1}\ge0,\,  x_{2}\leq0$ and Assumptions \ref{hypo1} and \ref{changed assumption for artifical cumulative dividend price} be valid. \hfill \break
(i) If $x_{1}x_{2}=0$, then for any contract $(A,C)$ such that $A \in \Classaptb $ we have, for every $t\in[0,T]$
\be \label{new1}
P^{c}_t (x_{2},-A,-C)\leq P^{h}_t (x_{1},A,C),  \quad \PTb-\aass
\ee
(ii) Let $r^{l}$ and $r^{b}$ be deterministic and satisfy $r^{l}_{t}<r^{b}_{t}$ for all $t\in[0,T]$. Then
\eqref{new1} holds for all contracts $(A,C)$ such that $A \in \Classaptb $ and all $t \in [0,T]$  if and only if $x_{1}x_{2}=0$.
\ep

\begin{proof}
(i) Assume that $x_{1}\ge0$ and  $x_{2}\leq0$. We will show that
\begin{align*}
\delta&:=\overline{g}(t,x_{2},\overline{Y}^{1},\overline{Z}^{1})-\overline{f}(t,x_{1},\overline{Y}^{1},\overline{Z}^{1}) \\
&\ge \max \, \big\{-(\rbb_t- \rll_t)x_{1}B_{t}^{l}+\sumik_{i=1}^d(r^{i,b}_t-\rbb_t)|\overline{Z}^{1,i}_tS^i_t|,
(\rbb_t- \rll_t )x_{2}B_{t}^{b}+\sumik_{i=1}^d(r^{i,b}_t-\rll_t)|\overline{Z}^{1,i}_tS^i_t| \big\}.
\end{align*}
Indeed, we have
\bde
\delta =x_{1}r_{t}^{l}B_{t}^{l}+x_{2}r_{t}^{b}B_{t}^{b} - \rll_t (\delta_{1}^{+}+\delta_{2}^{+}) +
\rbb_t (\delta_{1}^{-}+\delta_{2}^{-})
\ede
where
\bde
\delta_{1}:=-\overline{Y}_{t}^{1}-q(-\overline{Y}_{t}^{1})+x_2 B_{t}^{b}+\sumik_{i=1}^d(-\overline{Z}^{1,i}_tS^i_t)^{-},\quad \delta_{2}:=\overline{Y}_{t}^{1}+q(-\overline{Y}_{t}^{1})+x_1 B_{t}^{l}+\sumik_{i=1}^d(\overline{Z}^{1,i}_tS^i_t)^{-}.
\ede
From $\rll_{t}\leq\rbb_t$, it follows that
\bde
\delta\ge \sumik_{i=1}^dr^{i,b}_t|z^{i}S^i_t|+x_{1}r_{t}^{l}B_{t}^{l}+x_{2}r_{t}^{b}B_{t}^{b} - \rll_t (\delta_{1}+\delta_{2})
=(\rbb_t- \rll_t )x_{2}B_{t}^{b}+\sumik_{i=1}^d(r^{i,b}_t-\rll_t)|\overline{Z}^{1,i}_tS^i_t|
\ede
and
\bde
\delta\ge \sumik_{i=1}^dr^{i,b}_t|z^{i}S^i_t|+x_{1}r_{t}^{l}B_{t}^{l}+x_{2}r_{t}^{b}B_{t}^{b} - \rbb_t (\delta_{1}+\delta_{2})= -(\rbb_t- \rll_t)x_{1}B_{t}^{l}+\sumik_{i=1}^d(r^{i,b}_t-\rbb_t)|\overline{Z}^{1,i}_tS^i_t|.
\ede
We have thus shown that
\bde
\delta \ge \max \, \big\{-(\rbb_t- \rll_t)x_{1}B_{t}^{l}+\sumik_{i=1}^d(r^{i,b}_t-\rbb_t)|\overline{Z}^{1,i}_tS^i_t|,
(\rbb_t- \rll_t )x_{2}B_{t}^{b}+\sumik_{i=1}^d(r^{i,b}_t-\rll_t)|\overline{Z}^{1,i}_tS^i_t| \big\}.
\ede
If $x_{1}x_{2}=0$, then using $r^{i,b}_t\ge r^b_t\ge r^{l}_t$, it is easy to check that the right-hand side of the above inequality is non-negative. Hence $\delta\ge0$ and thus,  from the comparison theorem for BSDEs and Proposition \ref{opposite sign hedger collateral dependent Bergman model ex-dividend price}, we deduce that inequality \eqref{new1} holds for every $t\in[0,T]$. \vskip 5 pt
\noindent (ii) If $x_{1}x_{2} \neq 0$, then the example from the proof of Proposition 5.4 in \cite{NR2} gives a contract $(A,C)$ with $q\equiv0$, such that the inequality
\bde
P^{c}_{0} (x_{2},-A,-C) >  P^{h}_{0} (x_{1},A,C),  \quad \PTb-\aass
\ede
holds in the present framework, so that ${\cal R}^p_0(x_1,x_2)$ is non-empty almost surely.
\end{proof}

\brem\label{monotone contract}
If $x_1x_2<0$ then, from the above proposition, we know that for some contracts $(A,C)$, we have
$P^{c}_{\widehat{t}} (x_{2},-A,-C) >  P^{h}_{\widehat{t}} (x_{1},A,C)$
for some $\widehat{t}\in[0,T]$. As in \cite{NR2}, for some special contracts of $(A,C)$, the inequality $P^{c}_{t} (x_{2},-A,-C) \leq P^{h}_{t} (x_{1},A,C)$ holds for all $t\in[0,T]$.  We will discuss in next subsection.
\erem

\subsubsection{Contracts with Monotone Cash Flows}  \label{sec5.3}

We continue the study of the case where $x_1 \ge0$ and $x_2 \le 0$. Motivated by \cite{NR2}, we will show that for some special contracts $(A,C)$,  inequality \eqref{new1} holds for all $t\in[0,T]$.

\bhyp \label{assx}
The following conditions are satisfied by a contract  $(A,C)$: \hfill \break
(i) the process $A - A_0 $ is decreasing and belongs to  the class $\Classaptb $, \hfill \break
(ii) the collateral $C$ is given by \eqref{collateral 1} with the function $q$ satisfying $y+q(-y)\ge0$ for all $y\ge 0$.
\ehyp

Condition (ii) holds, for instance, when $q(y)=(1+\alpha_1)y^{+}-(1+\alpha_2)y^{-}$
for some haircut processes $\alpha_1, \alpha_2$ such that $\alpha_2\leq 0$ which means that when the hedger
posts collateral then the cash amount never exceeds the full collateral.
Indeed, $q$ is obviously uniformly Lipschitz continuous and $q(0)=0$. Moreover,  we have, for all $y\ge0$,
\[
y+q(-y)=y-(1+\alpha_2)y=-\alpha_2 y\ge 0.
\]
To emphasize the important role of the function $q$, we will sometimes write $P^{h}_t (x_1,A,q)$ and $P^{c}_t (x_1,-A,-q)$ instead of
$P^{h}_t (x_1,A,C)$ and $P^{c}_t (x_2,-A,-C)$, respectively.

\brem In the case of Bergman's model, we were unable to prove that the
range of fair bilateral prices is non-empty using the method employed to establish the next result. This shows once again
that the properties of prices depend on specific features of a market model at hand.
\erem

\bp \label{monotone contract partial netting model inequality proposition for positive negative initial wealth}
Let $x_{1}\ge0,\,  x_{2}\leq0$ and Assumptions \ref{hypo1} and \ref{changed assumption for artifical cumulative dividend price} be valid.
If a contract $(A,C)$ satisfies Assumption \ref{assx}, then the inequality $P^{c}_t (x_{2},-A,-q)\leq P^{h}_t (x_{1},A,q)$
holds for every $t\in[0,T]$.
\ep

\begin{proof}
We already know that the pair $(P^{h}_{t},\widetilde{Z}^{h,x_1}_t)$ solves BSDE
(\ref{partial netting main BSDE hedger 1}), whereas the pair $(P^{c}_{t},\widetilde{Z}^{c,x_2}_t)$ solves BSDE
(\ref{partial netting main BSDE counterparty 1}). Noticing that $\overline{f}(t,x_1,0,0)=0$ and $A-A_0$ is a decreasing process, from the comparison theorem for BSDEs, we obtain $P^{h}=\overline{Y}^1\ge0$. Therefore, from $x_1\ge0$ and
\[
y+q(-y)\ge0 \text{ for all }y\ge 0,
\]
we get
\bde
\begin{array}[c]{rl}
\overline{f}(t,x_1,\overline{Y}^1_t,\overline{Z}^1_t)&=\sum_{i=1}^d\overline{Z}^{i}_t\beta^{i}_{t}S_{t}^{i}
-\sumik_{i=1}^dr^{i,b}_t(\overline{Z}^{i}_tS^i_t)^{+}-x_1\rll_t\Blr_t-r_{t}^{c}q(-\overline{Y}^1_t)
\medskip\\
&\quad \mbox{} +r_{t}^{l}\Big(\overline{Y}^1_t+q(-\overline{Y}^1_t)+x_1B_{t}^{l}+\sumik_{i=1}^d(\overline{Z}^{i}_tS^i_t)^{-}\Big)\medskip\\
&=\sum_{i=1}^d\overline{Z}^{i}_t\beta^{i}_{t}S_{t}^{i}
-\sumik_{i=1}^dr^{i,b}_t(\overline{Z}^{i}_tS^i_t)^{+}-r_{t}^{c}q(-\overline{Y}^1_t)
\medskip\\
&\quad \mbox{} +r_{t}^{l}\Big(\overline{Y}^1_t+q(-\overline{Y}^1_t)+\sumik_{i=1}^d(\overline{Z}^{i}_tS^i_t)^{-}\Big).
\end{array}
\ede
Since
\bde
\begin{array}[c]{rl}
\overline{g}(t,x_2,y,z)\ge &\sum_{i=1}^dz^{i}\beta^{i}_{t}S_{t}^{i}+\sumik_{i=1}^dr^{i,b}_t(-z^{i}S^i_t)^{+}+x_2\rbb_t\Bbr_t-r_{t}^{c}q(-\overline{Y}^1_{t})\medskip\\
&\ \mbox{} -r_{t}^{b}\Big(-y-q(-\overline{Y}^1_{t})+x_2B_{t}^{b}+\sumik_{i=1}^d(-z^{i}S^i_t)^{-}\Big)\medskip\\
&=\sum_{i=1}^dz^{i}\beta^{i}_{t}S_{t}^{i}+\sumik_{i=1}^dr^{i,b}_t(-z^{i}S^i_t)^{+}-r_{t}^{c}q(-\overline{Y}^1_{t})\medskip\\
&\ \mbox{} -r_{t}^{b}\Big(-y-q(-\overline{Y}^1_{t})+\sumik_{i=1}^d(-z^{i}S^i_t)^{-}\Big),
\end{array}
\ede
we have that
\bde
\begin{array}[c]{rl}
&\overline{g}(t,x_1,\overline{Y}^1_t,\overline{Z}^1_t)-\overline{f}(t,x_1,\overline{Y}^1_t,\overline{Z}^1_t)\medskip\\
&\ge  \sumik_{i=1}^dr^{i,b}_t|\overline{Z}^{i}_tS^i_t|+(\rbb_t-\rll_t)\Big(\overline{Y}^1_t+q(-\overline{Y}^1_t)\Big)
-\rll_t\sumik_{i=1}^d(\overline{Z}^{i}_tS^i_t)^{-}-\rbb_t\sumik_{i=1}^d(-\overline{Z}^{i}_tS^i_t)^{-}\medskip\\
& =(\rbb_t-\rll_t)\Big(\overline{Y}^1_t+q(-\overline{Y}^1_t)\Big)
+\sumik_{i=1}^d(r^{i,b}_t-\rll_t)(\overline{Z}^{i}_tS^i_t)^{-}+\sumik_{i=1}^d(r^{i,b}_t-\rbb_t)(-\overline{Z}^{i}_tS^i_t)^{-}.
\end{array}
\ede
In view of the inequalities $r^{i,b}\ge r^b\ge r^l,\, \overline{Y}^1\ge0$ and $y+q(-y)\ge0$ for all $y\ge0$, we conclude that
\[
\overline{g}(t,x_1,\overline{Y}^1_t,\overline{Z}^1_t)-\overline{f}(t,x_1,\overline{Y}^1_t,\overline{Z}^1_t)\ge 0
\]
and thus the comparison theorem for BSDEs yields the desired inequality.
\end{proof}

\subsection{Price Independence of the Hedger's Initial Endowment}

Our next goal is to demonstrate that for a certain class of contracts  the hedger's price in the model with partial netting is independent of the initial endowment $x_1$. The financial interpretation of Proposition \ref{pro_new1} is that the hedger will never need to borrow
cash from the account $B^b$ for hedging purposes and thus the actual level of his non-negative initial endowment is
immaterial for his pricing problem. It is thus clear that a similar result will not hold when $x_1\leq0$.
By the same token, the independence property  will not hold in Bergman's model, in general, since in the latter model the funding of
positive positions in risky assets may require borrowing from the cash account $B^b$.

\bp \label{pro_new1}
Let $x_{1}\ge0$ and Assumptions \ref{hypo1} and \ref{changed assumption for artifical cumulative dividend price} be valid.
If a contract $(A,C)$  satisfies Assumption \ref{assx}, then the hedger's price $P^{h}_{t} (x_{1},A,q)$ is independent of $x_1$.
\ep

\begin{proof}
From Proposition \ref{opposite sign hedger collateral dependent partial netting model ex-dividend price}, we have
$P^h(x_1,A,q)= \overline{Y}^1$ where $(\overline{Y}^{1}, \overline{Z}^{1})$ is the unique solution to the BSDE
\bde 
\left\{ \begin{array}
[c]{l}
d\overline{Y}_t^1 = \overline{Z}^{1,\ast}_t \, d \wt S^{{\textrm{cld}}}_t
+\overline{f} \big(t, x_1,\overline{Y}^1_t, \overline{Z}^1_t \big)\, dt + dA_t, \medskip\\
\overline{Y}^1_T=0.
\end{array} \right.
\ede
 Since $\overline{f}(t,x_1,0,0)=0$ and $A_t-A_{0}$ is a decreasing process, from the comparison theorem for BSDEs, we obtain $\overline{Y}^1\ge0$. Therefore, using the inequalities $x_1\ge0$ and $y+q(-y)\ge0$ for all $y\ge 0$, we get
\bde
\begin{array}[c]{rl}
\overline{f}(t,x_1,\overline{Y}^1_t,\overline{Z}^1_t)&=\sum_{i=1}^d\overline{Z}^{i}_t\beta^{i}_{t}S_{t}^{i}
-\sumik_{i=1}^dr^{i,b}_t(\overline{Z}^{i}_tS^i_t)^{+}-x_1\rll_t\Blr_t-r_{t}^{c}q(-\overline{Y}^1_t)
\medskip\\
&\quad \mbox{}+r_{t}^{l}\Big(\overline{Y}^1_t+q(-\overline{Y}^1_t)+x_1B_{t}^{l}+\sumik_{i=1}^d(\overline{Z}^{i}_tS^i_t)^{-}\Big)\medskip\\
&=\sum_{i=1}^d\overline{Z}^{i}_t\beta^{i}_{t}S_{t}^{i}
-\sumik_{i=1}^dr^{i,b}_t(\overline{Z}^{i}_tS^i_t)^{+}-r_{t}^{c}q(-\overline{Y}^1_t)
\medskip\\
&\quad \mbox{} +r_{t}^{l}\Big(\overline{Y}^1_t+q(-\overline{Y}^1_t)+\sumik_{i=1}^d(\overline{Z}^{i}_tS^i_t)^{-}\Big)
\end{array}
\ede
where the last expression is independent of $x_1$. Consequently, the price $P^h_{t}(x_{1},A,q)= \overline{Y}^1_{t}$ is also independent of $x_1$.
\end{proof}

\brem
Suppose that $x_2\ge0$ and a contract $(A,C)$ is such that the process $A - A_0$ is increasing and belongs to $\Classaptb $. If
the collateral $C$, as seen from the perspective of the hedger, is given by  $C_t=q(V_t^c - V_t^0(x_2))$ where the function $q$ satisfies $-y+q(y)\geq 0$ for all $y\geq 0$, then the counterparty's price $P^{c}_{t} (x_{2},-A,-q)$ is independent of~$x_2$.

However, if we still work under the assumption of the hedger's collateral, the problem requires more attention, since the counterparty's price depends also on the hedger's initial endowment $x_1$. As shown in above proposition, for a contract $(A,C)$ satisfying Assumption \ref{assx}, the process $\overline{Y}^1$ is independent of $x_1$ so that, obviously, the price $P^{c}_{t} (x_{2},-A,-q)$ is independent of $x_1$, but it still may depend on $x_2$. It is not clear at this moment whether one can find some class of non-trivial contracts $(A,C)$ with the hedger's collateral $C$ given by \eqref{collateral 1} such that $P^{c}_{t} (x_{2},-A,-q)$ does not depend on $x_2$ (it may still depend on $x_1$).
\erem

\subsection{Positive Homogeneity of the Hedger's Price}

We consider once again the hedger's price and we show that it is positively homogeneous with respect to
the size of the contract and the non-negative initial endowment. Observe that this property is no longer true if only
the size of the contract, but not the level of the hedger's initial endowment, is inflated (or deflated) through
a non-negative scaling factor $\lambda $. Of course, this comment does not apply when the price of a contract is known to be independent of the hedger's initial endowment as is the case, for instance, under the assumptions of Proposition \ref{pro_new1}.

\bp \label{pro_new2}
Let $x_{1}\ge0$ and Assumptions \ref{hypo1} and  \ref{changed assumption for artifical cumulative dividend price} be valid.
For any contract $(A,C)$ such that $A-A_0\in\Classaptb $ and the function $q$ in equation \eqref{collateral 1} is positively homogeneous,
meaning that $q(\lambda y)=\lambda q(y)$ for all $\lambda\ge0$, then the hedger's price is positively homogeneous as well, specifically,
for all $\lambda\in\mathbb{R}_{+}$  and $t\in[0,T]$,
\be\label{homogeneous 2}
P^{h}_t (\lambda x_{1},\lambda A,q)=\lambda P^{h}_t (x_1,A,q), \quad \PTb-\aass
\ee
\ep

\begin{proof}
It is obvious that (\ref{homogeneous 2}) holds for $\lambda=0$. Now we suppose that $\lambda>0$.
From  Proposition \ref{opposite sign hedger collateral dependent partial netting model ex-dividend price}, we know that
$P^h(x_1,A,q)= \overline{Y}^1$ where $(\overline{Y}^{1}, \overline{Z}^{1})$ is the unique solution to the BSDE
\bde
\left\{ \begin{array}
[c]{l}
d\overline{Y}_t^1 = \overline{Z}^{1,\ast}_t \, d \wt S^{{\textrm{cld}}}_t
+\overline{f} \big(t, x_1,\overline{Y}^1_t, \overline{Z}^1_t \big)\, dt + dA_t, \medskip\\
\overline{Y}^1_T=0.
\end{array} \right.
\ede
Similarly, $P^h(\lambda x_1,\lambda A,q)= \widetilde{Y}^1$ where $(\widetilde{Y}^{1}, \widetilde{Z}^{1})$ is the unique solution to the BSDE
\bde
\left\{ \begin{array}
[c]{l}
d\widetilde{Y}_t^1 = \widetilde{Z}^{1,\ast}_t \, d \wt S^{{\textrm{cld}}}_t
+\overline{f} \big(t, \lambda x_1,  \widetilde{Y}^1_t, \widetilde{Z}^1_t \big)\, dt + \lambda \, dA_t, \medskip\\
\widetilde{Y}^1_T=0.
\end{array} \right.
\ede
Hence for $\overline{Y}:=\lambda \overline{Y}^1$ and $\overline{Z}=\lambda \overline{Z}^1$ we have
\bde
\left\{ \begin{array}
[c]{l}
d\overline{Y}_t = \overline{Z}^{\ast}_t \, d \wt S^{{\textrm{cld}}}_t
+\lambda\overline{f} \big(t, x_1,\lambda^{-1}\overline{Y}_t, \lambda^{-1}\overline{Z}_t \big)\, dt + \lambda \, dA_t, \medskip\\
\overline{Y}_T=0.
\end{array} \right.
\ede
To complete the proof, it is sufficient to show that for every $\lambda\in\mathbb{R}_{+}$
\[
\lambda\overline{f} \big(t, x_1,\lambda^{-1}y, \lambda^{-1}z \big)=\overline{f} \big(t, \lambda x_1,y, z \big).
\]
This can be checked easily using the property $q(\lambda y)=\lambda q(y)$ for every $\lambda\in\mathbb{R}_{+}$.
\end{proof}

If the collateral is given by $C_t=q(V_t^c - V_t^0(x_2))$, then the counterparty's price has the similar positive homogeneity property as in
Proposition \ref{pro_new2}. However, if the hedger's collateral is postulated, the study of the homogeneity property of the counterparty's price is slightly more complex, since the counterparty's price depends, in particular, on the hedger's initial endowment $x_1$.
In that case  $C_t=q(V_t^0(x_1)-V_t^h)$, which depends on $(x_1,A)$, so we shall write $C_t=C^{x_1,A}_t$.

\bp  \label{pro_new3}
Let $x_{2}\leq0$ and Assumptions \ref{hypo1} and  \ref{changed assumption for artifical cumulative dividend price} be valid.
For any contract $(A,C)$ such that $A-A_0\in\Classaptb $ and the function $q$ in equation \eqref{collateral 1} is positively homogeneous
we have, for all $\lambda\in\mathbb{R}_{+}$  and $t\in[0,T]$,
\bde 
P^{c}_t (\lambda x_{2},-\lambda A,C^{\lambda x_1,\lambda A})=\lambda P^{c}_t (x_2,-A, C^{x_1,A}), \quad \PTb-\aass
\ede
\ep

\begin{proof}
Similarly as in the proof of Proposition \ref{pro_new2}, it is now sufficient to show that
\[
\lambda\overline{g} \big(t, x_2,\lambda^{-1}y, \lambda^{-1}z \big)=\overline{g} \big(t, \lambda x_2,y, z \big)
\]
where the function $\overline{g} $ is given in Proposition \ref{opposite sign hedger collateral dependent partial netting model ex-dividend price}. Since $q(\lambda y)=\lambda q(y)$ and $\widetilde{Y}^1=\lambda \overline{Y}^1$ for $\lambda\ge0$ (see Proposition \ref{pro_new2}),
it is easy to complete the proof.
\end{proof}

\brem
Results similar to Propositions \ref{pro_new2} and \ref{pro_new3} are also valid when the initial endowments satisfy
$x_1\le0$ and $x_2\ge0$. Moreover, by combining the results of the last two sections, we can find
a class of contracts with prices that are independent of initial endowments and positively homogeneous.
Analogous price homogeneity properties can also be established for Bergman's model. The proofs are fairly similar to those for the model
with partial netting  and thus they are not presented here.
\erem

\section{Model with Partial Netting and Negotiated Collateral} \label{sec6}

In the final section, we continue the analysis of the model with partial netting by studying the case where the collateral
amount $C$ is negotiated between the counterparties, in the sense that it depends on both the hedger's value $V^{h}:=V(x_{1},\varphi,A,C)$ and the counterparty's value $V^{c}:=V(x_{2},\widetilde{\varphi},-A,-C)$. As in Section \ref{sec4}, we postulate that the collateral satisfies Assumption \ref{hypo2}. Recall that in that case we have $P^{h}(x_{1},A,C)= P^{h}(x_{1},x_2,A,C)$ and $P^{c}(x_2,-A,-C)= P^{c}(x_{1},x_2,-A,-C)$, meaning that the two prices depend on the vector $(x_1,x_2)$ of initial endowments.

\subsection{Initial Endowments of Equal Signs} \label{sec6.2}

The following result gives fully-coupled pricing BSDEs for both parties under the assumption that their initial
endowments are non-negative.

\bp \label{symmetric collateral dependent partial netting model ex-dividend price}
Let $x_1\ge0,\, x_2\ge0$ and  Assumptions  \ref{changed assumption for lending cumulative dividend price} and \ref{hypo2} be valid. For any contract $(A,C)$ such that $A \in \Classapl $ we have $(P^{h},P^{c})^\ast=Y$ where $(Y,Z)$ solves the two-dimensional fully-coupled BSDE
\be \label{partial netting main two dimensioanl BSDE}
\left\{ \begin{array}
[c]{l}
dY_t = Z^{\ast}_t \, d \wt S^{l,{\textrm{cld}}}_t
+g \big(t,Y_t, Z_t \big)\, dt + d\overline{A}_t, \medskip\\
Y_T=0,
\end{array} \right.
\ee
where $g=(g^1,g^2)^\ast,\, \overline{A}=(A,A)^\ast$ and for all $y=(y_1,y_2)^\ast\in\mathbb{R}^{2},\, z=(z_1,z_2)\in\mathbb{R}^{d\times 2}$,
\be \label{partial netting generator 1}
\begin{array}[c]{rl}
g^{1}(t,y,z)=&r_{t}^{l}(B_t^{l})^{-1}z^{\ast}_1 S_t -(B_{t}^{l})^{-1}\sumik_{i=1}^dr^{i,b}_t(z^{i}_1S^i_t)^{+}-x_1 B_{t}^{l}r_{t}^{l}-r_{t}^{c}\Cnq (-y_1,-y_2)\medskip\\
&\mbox{}+r_{t}^{l}\Big(y_1+\Cnq (-y_1,-y_2)+x_1B_{t}^{l}+(B_{t}^{l})^{-1}\sumik_{i=1}^d(z^{i}_1S^i_t)^{-}\Big)^+\medskip\\
&\mbox{}-r_{t}^{b}\Big(y_1+\Cnq (-y_1,-y_2)+x_1B_{t}^{l}+(B_{t}^{l})^{-1}\sumik_{i=1}^d(z^{i}_1S^i_t)^{-}\Big)^-
\end{array}
\ee
and
\be \label{partial netting generator 2}
\begin{array}[c]{rl}
g^{2}(t,y,z)=&r_{t}^{l}(B_t^{l})^{-1}z^{\ast}_2 S_t+ (B_{t}^{l})^{-1}\sumik_{i=1}^dr^{i,b}_t(-z^{i}_2S^i_t)^{+}+x_2 B_{t}^{l}r_{t}^{l}-r_{t}^{c}\Cnq (-y_1,-y_2)\medskip\\
&\mbox{}-r_{t}^{l}\Big(-y_2-\Cnq (-y_1,-y_2)+x_2B_{t}^{l}+(B_{t}^{l})^{-1}\sumik_{i=1}^d(-z^{i}_2S^i_t)^{-}\Big)^+\medskip\\
&\mbox{}+r_{t}^{b}\Big(-y_2-\Cnq (-y_1,-y_2)+x_2B_{t}^{l}+(B_{t}^{l})^{-1}\sumik_{i=1}^d(-z^{i}_2S^i_t)^{-}\Big)^-.
\end{array}
\ee
\ep

In the remainder of this section, we work under Assumption \ref{assdif} and we study the valuation and hedging of a European contingent claim $(H_T,C)$. We note that BSDE (\ref{partial netting main two dimensioanl BSDE}) becomes
\be \label{partial netting main two dimensioanl Brownian BSDE with general A}
\left\{ \begin{array}
[c]{l}
dY_t = Z_t\sigma (t, S_{t})\, dW_t
+(g \big(t,Y_t, Z_t \big)+\sigma (t, S_{t}) a_tZ_t)\, dt + d\overline{A}_t, \medskip\\
Y_T=0 ,
\end{array} \right.
\ee
where the process $a$ is given by \eqref{defaa}.
As in Section \ref{sec4.3}, it suffices to examine the following BSDE
\bde
\left\{ \begin{array}
[c]{l}
dY_t = Z_t\, d\wtWl_t
+g \big(t,Y_t, (\sigma(t, S_{t}))^{-1}Z_t \big)\, dt, \medskip\\
Y_T=(-H_T,-H_T)^{\ast}.
\end{array} \right.
\ede
We are now in a position to study the range of fair bilateral prices at time $t$ for a collateralized European claim.

\bp \label{general partial netting model inequality proposition for both positive initial}
Let $x_{1}\ge0,\, x_{2}\ge0$ and Assumptions \ref{changed assumption for lending cumulative dividend price}, \ref{assdif}, \ref{hypo2} and \ref{assumption for stock coefficients} be valid.
For any collateralized European claim $(H_T,C)$ where $H_T \in L^{2}(\Omega,\mathcal{F}_T, \PT^l )$ we have, for every $t\in[0,T]$,
\bde
P^{c}_t (x_{2},-H_T,-C)\leq P^{h}_t (x_{1},H_T,C),  \quad \PT^l-\aass
\ede
\ep

\begin{proof}
Let $\sigma^{-1}_t:=(\sigma(t, S_{t}))^{-1}$.
It is sufficient to check that the functions
\[
h^{1}(t,y_1,y_2,z_1,z_2):=-g^{1}\big(t,y_1,y_2,\sigma^{-1}_tz_1,\sigma^{-1}_tz_2\big)
\]
and
\[
h^{2}(t,y^1,y_2,z_1,z_2):=-g^{2}\big(t,y_1,y_2,\sigma^{-1}_tz_1,\sigma^{-1}_tz_2\big)
\]
 where $g^{1}$ and $g^{2}$ are given by (\ref{partial netting generator 1}) and (\ref{partial netting generator 2}) with $d=1$,
 satisfy Assumption  \ref{assumption for Brownian BSDE generator} and condition (\ref{BSVP condition}). It is easy to check that  Assumption  \ref{assumption for Brownian BSDE generator} holds. We will check that condition (\ref{BSVP condition}) is satisfied as well. We set
\[
\delta_1:=y_1^++y_2+\Cnq (-y_1^+-y_2,-y_2)+x_1B_{t}^{l}+(B_{t}^{l})^{-1}\sigma^{-1}_t((z_1+z_2) S_t)^{-}
\]
and
\[
\delta_2:=-y_2-\Cnq (-y_1^+-y_2,-y_2)+x_2B_{t}^{l}+(B_{t}^{l})^{-1}\sigma^{-1}_t(-z_2 S_t)^{-}.
\]
Then
\[
\begin{array}[c]{rl}
&h^{1}(t,y_1^++y_2,y_2,z_1+z_2,z_2)-h^{2}(t,y_1^++y_2,y_2,z_1+z_2,z_2)\medskip\\
&=-g^{1}(t,y_1^++y_2,y_2,\sigma^{-1}_t(z_1+z_2),\sigma^{-1}_tz_2)+g^{2}(t,y_1^++y_2,y_2,\sigma^{-1}_t(z_1+z_2),\sigma^{-1}_tz_2)\medskip\\
&=-r_{t}^{l}(B_t^{l})^{-1}\sigma^{-1}_tz_1 S_t +(B_{t}^{l})^{-1}r^{1,b}_t(\sigma^{-1}_t(z_1+z_2)S_t)^{+}+(B_{t}^{l})^{-1}r^{1,b}_t(-\sigma^{-1}_tz_2S_t)^{+}\medskip\\
&\mbox{}\ \ +(x_1+x_2) B_{t}^{l}r_{t}^{l}-r_{t}^{l}(\delta_1^++\delta_2^+)+r_{t}^{b}(\delta_1^-+\delta_2^-).
\end{array}
\]
Since $r_{t}^{l}\leq r_{t}^{b}$, we have
\[
\begin{array}[c]{rl}
& r_{t}^{l}(\delta_1^++\delta_2^+)-r_{t}^{b}(\delta_1^-+\delta_2^-) \leq  r_{t}^{l}(\delta_1+\delta_2)\medskip\\
&= r_{t}^{l}y_1^++ (x_1+x_2) B_{t}^{l}r_{t}^{l}+r_{t}^{l}(B_t^{l})^{-1}((\sigma^{-1}_t(z_1+z_2) S_t)^{-}+(-\sigma^{-1}_tz_2S_t)^{-}).
\end{array}
\]
Thus, using $r^{1,b}_t \ge r^l_t $, we obtain
\[
\begin{array}[c]{rl}
&h^{1}(t,y_1^++y_2,y_2,z_1+z_2,z_2)-h^{2}(t,y_1^++y_2,y_2,z_1+z_2,z_2)\medskip\\
&\ge-r_{t}^{l}y_1^+-r_{t}^{l}(B_t^{l})^{-1}\sigma^{-1}_tz_1 S_t +(B_{t}^{l})^{-1}r^{1,b}_t(\sigma^{-1}_t(z_1+z_2)S_t)^{+}+(B_{t}^{l})^{-1}r^{1,b}_t(-\sigma^{-1}_tz_2S_t)^{+}\medskip\\
&\mbox{}\ \ -r_{t}^{l}(B_t^{l})^{-1}((\sigma^{-1}_t(z_1+z_2) S_t)^{-}+(-\sigma^{-1}_tz_2S_t)^{-})\medskip\\
&= -r_{t}^{l}y_1^+ +(B_{t}^{l})^{-1}(r^{1,b}_t-r^l_t)(\sigma^{-1}_t(z_1+z_2)S_t)^{+}+(B_{t}^{l})^{-1}(r^{1,b}_t-r^l_t)(-\sigma^{-1}_tz_2S_t)^{+}\ge  -r_{t}^{l}y_1^+ .
\end{array}
\]
Using similar arguments as in the proof of Proposition \ref{general Bergman model inequality proposition for both positive initial wealth}, we conclude  that  (\ref{BSVP condition}) holds.
\end{proof}

\subsection{Initial Endowments of Opposite Signs}  \label{sec6.3}

We conclude the paper by studying the case of initial endowments of opposite signs.

\bp \label{opposite sign symmetric collateral dependent partial netting model ex-dividend price}
Let $x_{1}\ge0,\, x_{2}\leq0$ and Assumptions \ref{changed assumption for artifical cumulative dividend price} and \ref{hypo2}  be valid.
For any contract $(A,C)$ such that $A \in \Classaptb $ we have $(P^{h},P^{c})^\ast=\widehat{Y}$ where $(\widehat{Y},\widehat{Z})$ solves the two-dimensional fully-coupled BSDE
\be \label{partial netting main two dimensioanl BSDE 2}
\left\{ \begin{array}
[c]{l}
d\widehat{Y}_t = \widehat{Z}^{\ast}_t \, d \wt S^{{\textrm{cld}}}_t
+\widehat{g}\big(t,\widehat{Y}_t,\widehat{ Z}_t \big)\, dt + d\overline{A}_t, \medskip\\
\widehat{Y}_T=0,
\end{array} \right.
\ee
where $\widehat{g}=(\widehat{g}^1,\widehat{g}^2)^\ast,\, \overline{A}=(A,A)^\ast$ and for all $y=(y_1,y_2)^\ast\in\mathbb{R}^{2},\, z=(z_1,z_2)\in\mathbb{R}^{d\times 2}$,
\be \label{partial netting generator 3}
\begin{array}[c]{rl}
\widehat{g}^{1}(t,y,z)=&\sum_{i=1}^dz^{i}_1\beta^{i}_{t}S_{t}^{i} -x_1 B_{t}^{l}r_{t}^{l}-r_{t}^{c}\Cnq (-y_1,-y_2)\medskip\\
&\mbox{} +r_{t}^{l}\Big(y_1+\Cnq (-y_1,-y_2)+x_1B_{t}^{l}-z^{\ast}_1 S_t\Big)^+\medskip\\
&\mbox{}-r_{t}^{b}\Big(y_1+\Cnq (-y_1,-y_2)+x_1B_{t}^{l}-z^{\ast}_1 S_t\Big)^-
\end{array}
\ee
and
\be \label{partial netting generator 4}
\begin{array}[c]{rl}
\widehat{g}^{2}(t,y,z)=&\sum_{i=1}^dz^{i}_1\beta^{i}_{t}S_{t}^{i} +(B_{t}^{l})^{-1}\sumik_{i=1}^dr^{i,b}_t(-z^{i}_2S^i_t)^{+}+x_2 B_{t}^{b}r_{t}^{b}-r_{t}^{c}\Cnq (-y_1,-y_2)\medskip\\
&\mbox{}-r_{t}^{l}\Big(-y_2-\Cnq (-y_1,-y_2)+x_2B_{t}^{b}+(B_{t}^{l})^{-1}\sumik_{i=1}^d(z^{i}_2S^i_t)^{-}\Big)^+\medskip\\
&\mbox{}+r_{t}^{b}\Big(-y_2-\Cnq (-y_1,-y_2)+x_2B_{t}^{b}+(B_{t}^{l})^{-1}\sumik_{i=1}^d(z^{i}_2S^i_t)^{-}\Big)^-.
\end{array}
\ee
\ep
Let Assumptions \ref{assdif} and \ref{changed assumption for artifical cumulative dividend price} be satisfied. Then
\bde
d\wt S^{\textrm{cld}}_t =
\big( \mu(t, S_{t})+\kappa(t, S_{t})-\beta_t S_{t}\big)\, dt + \sigma (t, S_{t})\, dW_t
\ede
and BSDE (\ref{partial netting main two dimensioanl BSDE 2}) becomes
\bde
\left\{ \begin{array}
[c]{l}
dY_t = Z_t\sigma (t, S_{t})\, dW_t
+(\widehat{g }\big(t,Y_t, Z_t \big)+\sigma (t, S_{t}) b_tZ_t)\, dt + d\overline{A}_t , \medskip\\
Y_T=0 ,
\end{array} \right.
\ede
where the process $b$ is given by \eqref{defbb}.
As in Section \ref{sec4.3}, it suffices to examine the following BSDE
\bde
\left\{ \begin{array}
[c]{l}
dY_t = Z_t\, d\wtWb_t +\widehat{g} \big(t,Y_t, (\sigma(t, S_{t}))^{-1}Z_t \big)\, dt, \medskip\\
Y_T=(-H_T,-H_T)^{\ast},
\end{array} \right.
\ede
where $\wtWb$ is a Brownian motion under an equivalent probability measure $\PTb $.

\bp \label{general partial netting model inequality proposition for opposite initial wealth}
Let $x_{1}\ge0,\, x_{2}\leq 0$ be such that $x_1x_2=0$. If Assumptions  \ref{assdif}, \ref{changed assumption for artifical cumulative dividend price}, \ref{hypo2}  and \ref{assumption 2 for stock coefficients}
are met, then for any collateralized European claim $(H_T,C)$ such that $H_T \in L^{2}(\Omega,\mathcal{F}_T, \PTb )$ we have, for every $t\in[0,T]$,
\bde
P^{c}_t (x_{2},-H_T ,-C)\leq P^{h}_t (x_{1},H_T,C),  \quad \PTb-\aass
\ede
\ep

\begin{proof}
We write, as usual, $\sigma^{-1}_t:=(\sigma(t, S_{t}))^{-1}$. It is sufficient to check that the functions
\[
h^{1}(t,y_1,y_2,z_1,z_2):=-\widehat{g}^{1}\big(t,y_1,y_2,\sigma^{-1}_tz_1,\sigma^{-1}_tz_2\big)
\]
and
\[
h^{2}(t,y^1,y_2,z_1,z_2):=-\widehat{g}^{2}\big(t,y_1,y_2,\sigma^{-1}_tz_1,\sigma^{-1}_tz_2 \big)
\]
where $\widehat{g}^{1}$ and $\widehat{g}^{2}$ are given by (\ref{generator 3}) and (\ref{generator 4}), respectively,
satisfy Assumption  \ref{assumption for Brownian BSDE generator} and condition (\ref{BSVP condition}).
This is similar to the proof of Proposition \ref{general Bergman model inequality proposition for opposite initial wealth}, using $x_1x_2=0$ and the same computations as in the proof of Proposition \ref{general partial netting model inequality proposition for both positive initial}. The details are left to the reader.
\end{proof}

\noindent {\bf Acknowledgement.}

The research of Tianyang Nie and Marek Rutkowski was supported under Australian Research
Council's Discovery Projects funding scheme (DP120100895).



\end{document}